\documentclass[twocolumn]{aastex63}

\usepackage{graphicx, amsmath}
\usepackage[caption=false]{subfig}
\usepackage{hyperref}
\usepackage{cleveref}
\usepackage{xspace}
\usepackage{rotating}
\usepackage{comment}
\usepackage{array}

\newcommand{\tcell}[2]{\parbox[t]{#1}{\raggedright\arraybackslash #2}}

\newcommand{\lya}{Ly$\alpha$\xspace}

\newcommand{\heii}{\ion{He}{2}~$\lambda1640$\xspace}
\newcommand{\nv}{\ion{N}{5}\xspace}
\newcommand{\civ}{\ion{C}{4}\xspace}
\newcommand{\oiii}{\ion{O}{3}]\xspace}
\newcommand{\OII}{[\ion{O}{2}]\xspace}
\newcommand{\OIII}{[\ion{O}{3}]\xspace}
\newcommand{\Hb}{H$\beta$\xspace}
\newcommand{\Ha}{H$\alpha$\xspace}
\newcommand{\kms}{\ensuremath{\mathrm{km}\,\mathrm{s}^{-1}}\xspace}

\newcommand{\mheii}{\ion{He}{2}}

\graphicspath{{./}{figures/}}

\begin{document}

\title{Population~III Host Candidates at $\boldsymbol {z\sim2}$: \\ Strong He~\textsc{II} $\boldsymbol{\lambda}\mathbf{1640}$ and Absent UV Metal Lines in HETDEX Ly$\boldsymbol\alpha$ Emitters}

\author[0009-0003-1893-9526]{Mahan Mirza Khanlari}
\altaffiliation{Corresponding author: mahanmkh@utexas.edu}
\affiliation{Department of Astronomy, The University of Texas at Austin, 2515 Speedway Boulevard, Stop C1400, Austin, TX 78712, USA}

\author[0000-0002-8925-9769]{Dustin Davis}
\affiliation{Department of Astronomy, The University of Texas at Austin, 2515 Speedway Boulevard, Stop C1400, Austin, TX 78712, USA}

\author[0000-0002-8433-8185]{Karl Gebhardt}
\affiliation{Department of Astronomy, The University of Texas at Austin, 2515 Speedway Boulevard, Stop C1400, Austin, TX 78712, USA}

\author[0000-0003-4512-8705]{Tiger Yu-Yang Hsiao}
\affiliation{Department of Astronomy, The University of Texas at Austin, 2515 Speedway Boulevard, Stop C1400, Austin, TX 78712, USA}
\affiliation{Cosmic Frontier Center, The University of Texas at Austin, Austin, TX 78712, USA}

\author[0000-0002-8984-0465]{Julian B. Muñoz}
\affiliation{Department of Astronomy, The University of Texas at Austin, 2515 Speedway Boulevard, Stop C1400, Austin, TX 78712, USA}
\affiliation{Cosmic Frontier Center, The University of Texas at Austin, Austin, TX 78712, USA}

\author[0000-0003-2237-0777]{Alessandra Venditti}
\affiliation{Department of Astronomy, The University of Texas at Austin, 2515 Speedway Boulevard, Stop C1400, Austin, TX 78712, USA}
\affiliation{Cosmic Frontier Center, The University of Texas at Austin, Austin, TX 78712, USA}

\author[0000-0002-2307-0146]{Erin {Mentuch Cooper}}
\affiliation{Department of Astronomy, The University of Texas at Austin, 2515 Speedway Boulevard, Stop C1400, Austin, TX 78712, USA}

\author[0000-0002-1328-0211]{Robin Ciardullo}
\affiliation{Department of Astronomy \& Astrophysics, The Pennsylvania State University, University Park, PA 16802, USA}
\affiliation{Institute for Gravitation and the Cosmos, The Pennsylvania State University, University Park, PA 16802, USA}

\author[0000-0002-0212-4563]{Olivia Curtis}
\affiliation{Department of Astronomy \& Astrophysics, The Pennsylvania State University, University Park, PA 16802, USA}
\affiliation{Institute for Gravitation and the Cosmos, The Pennsylvania State University, University Park, PA 16802, USA}

\author[0000-0003-2575-0652]{Daniel J. Farrow}
\affiliation{E. A. Milne Centre for Astrophysics
University of Hull, Cottingham Road, Hull, HU6 7RX, UK}
\affiliation{Centre of Excellence for Data Science,
Artificial Intelligence \& Modelling (DAIM),
University of Hull, Cottingham Road, Hull, HU6 7RX, UK}

\author[0000-0001-8519-1130]{Steven L. Finkelstein}
\affiliation{Department of Astronomy, The University of Texas at Austin, Austin, TX, USA}
\affiliation{Cosmic Frontier Center, The University of Texas at Austin, Austin, TX, USA}

\author[0000-0001-6842-2371]{Caryl Gronwall}
\affiliation{Department of Astronomy \& Astrophysics, The Pennsylvania
State University, University Park, PA 16802, USA}
\affiliation{Institute for Gravitation and the Cosmos, The Pennsylvania State University, University Park, PA 16802, USA}

\author[0000-0001-6717-7685]{Gary J. Hill} 
\affiliation{McDonald Observatory, The University of Texas at Austin, 2515 Speedway Boulevard, Stop C1402, Austin, TX 78712, USA}
\affiliation{Department of Astronomy, The University of Texas at Austin, 2515 Speedway Boulevard, Stop C1400, Austin, TX 78712, USA}

\author[0000-0002-0417-1494]{Wolfram Kollatschny}
\affiliation{Institut f\"ur Astrophysik und Geophysik, Universit\"at G\"ottingen, Friedrich-Hund-Platz 1, 37077 G\"ottingen, Germany}

\author[0000-0001-5561-2010]{Chenxu Liu}
\affiliation{South-Western Institute for Astronomy Research, Key Laboratory of Survey Science of Yunnan Province, Yunnan University, Kunming, Yunnan 650500, People's Republic of China}

\author[0000-0003-3823-8279]{Shiro Mukae}
\affiliation{Department of Physics, School of Advanced Science and Engineering, Faculty of Science and Engineering, Waseda University, 3-4-1 Okubo, Shinjuku, Tokyo 169-8555, Japan}
\affiliation{MIRAI Technology Institute, Shiseido Co., Ltd., 1-2-11, Takashima, Nishi-ku, Yokohama, Kanagawa, 222-0011, Japan}

\author[0000-0002-6186-5476]{Shun Saito}
\affiliation{Institute for Multi-messenger Astrophysics and Cosmology, Department of Physics, Missouri University of Science and Technology, 1315 N. Pine St., Rolla MO 65409, USA}
\affiliation{Kavli Institute for the Physics and Mathematics of the Universe (WPI), The University of Tokyo Institutes for Advanced Study (UTIAS), The University of Tokyo, Chiba 277-8583, Japan}

\author[0000-0001-7240-7449]{Donald P. Schneider}
\affiliation{Department of Astronomy \& Astrophysics, The Pennsylvania State University, University Park, PA 16802, USA}
\affiliation{Institute for Gravitation and the Cosmos, The Pennsylvania State University, University Park, PA 16802, USA}

\begin{abstract}
Population~III (Pop~III) stars are expected to produce extremely hard ionizing spectra, yet direct evidence for their signature --- strong narrow \heii\ emission with weak or absent metal lines --- remains elusive. Although Pop~III formation peaks at $z\gtrsim10$, models predict that nearly pristine gas pockets can survive to later times, making intermediate-redshift searches a probe of metal mixing across cosmic time. We search 109,545 high-confidence Ly$\alpha$-emitting galaxies at $1.9<z<2.3$ in the Hobby--Eberly Telescope Dark Energy Experiment's database for systems with strong He~\textsc{ii} $\lambda1640$ and no detected UV metal lines, recovering eight candidates and a first-order, tentative comoving number density of ${\sim}30~{\rm Gpc}^{-3}$. The eight-object stack yields \mheii\ EW$_{\rm rest}=28.7\pm6.7$~\AA, \mheii/Ly$\alpha=0.354\pm0.094$, and a resolution-corrected \mheii\ FWHM $=449\pm105~\kms$. We find no significant \nv, \civ, or \oiii\ emission; using the \mheii\ width to set the scale, we derive $3\sigma$ upper limits of \nv/\mheii\ $<0.167$, \civ/\mheii\ $<0.128$, and \oiii/\mheii\ $<0.140$. This combination of strong \mheii, elevated \mheii/Ly$\alpha$, moderate line width, and weak metal lines is difficult to reproduce with metal-enriched stellar populations, AGN narrow-line regions, shocks, or classical Wolf--Rayet features. We interpret the sample as candidates for very metal-poor or Pop~III-like ionizing sources at $z\sim2$. Definitive confirmation requires deeper rest-UV spectroscopy to measure \mheii\ and metal-line limits in individual objects, followed by rest-optical spectroscopy of H$\beta$, \OIII\ $\lambda4959,\lambda5007$, \OII\ $\lambda3727$, and H$\alpha$ to constrain gas-phase metallicity, ionization conditions, and AGN activity.
\end{abstract}


\section{Introduction}
\label{sec:intro}

Population~III, or Pop~III, stars are the first generation of stars expected to form from metal-free or nearly metal-free gas. They deliver the first heavy elements and ionizing photons into the universe, and initiate chemical feedback, which drives the transition from pristine gas to chemically enriched galaxies \citep[e.g.,][]{bromm_2013,jaacks_2018,klessen_glover_2023}. Identifying credible signatures of these populations remains one of the most challenging issues in observational astrophysics. Pop~III star-forming episodes are short-lived, the host systems are intrinsically faint, and even modest metal enrichment erases the cleanest spectral diagnostics of pristine gas \citep[e.g.,][]{schaerer_2002,schaerer_2003,raiter_2010,katz_2023}. A useful search for newly formed Pop~III stars thus requires diagnostics that simultaneously probe hard ionizing radiation and weak metal enrichment.

Nebular \heii emission is the most direct rest-frame UV probe of hard ionizing spectra. Producing \heii recombination photons requires ionizing radiation above 54.4~eV, a threshold that normal, typically low-mass, metal-enriched stellar populations generally produce few such photons. Very massive \citep[e.g.,][]{stacy_2016,hirano_2014}, metal-poor, or metal-free stars can generate much harder spectra \citep[e.g.,][]{tumlinson_shull_2000,bromm_kudritzki_loeb_2001,oh_2001,schaerer_2003,raiter_2010}.

\heii alone, however, is not a clean diagnostic. For instance, AGN can produce strong \heii and often possess high-ionization metal lines, especially \civ $\lambda 1550$ and \nv $\lambda 1240$ \citep[e.g.,][]{baldwin_1978,feltre_2016}. Fast shocks can generate hard radiation fields and UV line emission, and can also produce [\ion{O}{1}] $\lambda6300$, with line ratios and kinematics that depend on shock velocity, density, magnetic field, and metallicity \citep[e.g.,][]{allen_2008}. Wolf--Rayet stars, stripped stars, X-ray binaries, very massive stars, and binary stellar populations can also contribute to \heii under some conditions \citep[e.g.,][]{shirazi_brinchmann_2012,senchyna_2020}.

The key diagnostic is therefore the full line pattern. In the idealized case, Pop~III-like ionizing sources produce strong \heii and high \heii/Ly$\alpha$ while leaving metal lines weak or undetected \citep[e.g.,][]{schaerer_2003,nakajima_maiolino_2022}. In practice, this clean signature traces only a short-lived phase: self-enrichment from the Pop~III stars themselves, or contamination from surrounding gas, will produce metal-line emission even while Pop~III massive stars are still alive \citep[e.g.,][]{cleri_2023,rusta_2025}. Moreover, Pop~III star formation may occur in metal-free pockets of a metal-enriched ISM, and therefore show UV metal lines from the surrounding gas \citep[e.g.,][]{venditti_2023,venditti_2026}. A system with simultaneously strong \heii and no detectable UV metal-line emission remains extremely difficult to reproduce with standard metal-enriched contaminants, all of which deposit significant flux in at least one of \nv, \civ, or \oiii.

At $z\sim2$, pristine star formation would be surprising, but models predict that very low-metallicity star formation can persist to late cosmic times in regions where metal transport and mixing have been inefficient. Simulations and semi-analytic models reveal that pockets of nearly pristine gas can survive well beyond the first episodes of star-formation \citep{tornatore_2007,johnson_2013,sarmento_2018,venditti_2023}, and some studies suggest that inefficiently mixed primordial gas at $z\simeq3$--4 could produce hard rest-frame UV signatures including detectable \heii emission \citep{jimenez_haiman_2006}. These results do not make a Pop~III interpretation at $z\sim2$ probable, but they establish that rare-object searches for systems with Pop~III-like spectral signatures are physically justified at these redshifts. Even upper limits on Pop~III activity at $z\sim2$ constrain metal transport across cosmic time and test models of inhomogeneous enrichment. 

Observational studies of more general \heii-emitters at $z\sim2$--5 show that such systems span a broad range of rest-UV line ratios and inferred stellar-population properties \citep[e.g.,][]{nanayakkara_2019,nanayakkara_2021,saxena_2020}. In the local universe, the CLASSY survey \citep{berg_2022} offers a complementary benchmark. Nearby UV-bright star-forming galaxies exhibit rest-frame UV emission spanning \civ, \heii, and \oiii, demonstrating that \heii is not unique to primordial stellar populations and must therefore be interpreted alongside the accompanying metal-line pattern \citep{mingozzi_2022}. For Pop~III or extremely metal-poor candidates, previous searches have instead consistently relied on \heii together with the absence or weakness of metal-line emission, using Ly$\alpha$, \heii, Balmer lines, and metal-line non-detections rather than \heii alone \citep[e.g.,][]{nagao_2008,prescott_2009,fujimoto_2025,morishita_2025,hsiao_2025, venditti_2026,maiolino_2024,maiolino_2026}. This motivates the multi-line selection logic used here, which is designed to separate the rare very metal-poor class from the more common contaminants listed above.

Historically, \lya-selected galaxies were among the earliest proposed routes to finding young, chemically primitive systems \citep{partridge_peebles_1967,pritchet_1994}. Modern surveys have shown that Ly$\alpha$-emitting galaxies (LAEs) are not generally pristine, but they are often low-mass, low-dust, actively star-forming galaxies with sub-solar metallicities \citep[e.g.,][]{gawiser_2007,finkelstein_2011,nakajima_2013}. These properties make large LAE surveys a natural place to search for rare very metal-poor or Pop~III-like candidates.

The Hobby--Eberly Telescope Dark Energy Experiment \citep[HETDEX;][]{gebhardt_hobby-eberly_2021} is well matched to this search in three respects. First, it is an untargeted spectroscopic survey, so emission-line sources are selected from spectra with no imaging preselection, eliminating a systematic bias against UV-faint objects. Second, its $90~\mathrm{deg}^2$ footprint and catalog of more than $10^6$ LAEs provide the statistics needed to recover intrinsically rare populations. Third, at $1.9<z<2.3$, the Visible Integral-field Replicable Unit Spectrograph (VIRUS) ($3500$--$5500$~\AA) simultaneously covers Ly$\alpha$, \heii, \nv, \civ, and \oiii, so both the candidate signal and the key contaminant diagnostics are measured in the same spectra. The goal in this paper is not to collect every \heii emitter, but to isolate the rare subset with strong \heii, a high \heii/Ly$\alpha$ ratio, and weak or absent UV metal-line emission.

The remainder of the paper is organized as follows. Section~\ref{sec:data} describes the HETDEX data and LAE parent sample. Section~\ref{sec:candidate_selection} describes the \heii search, vetting, and final sample selection. Section~\ref{sec:extraction_stacking_measurements} details the spectral extraction, stacking, line measurements, robustness tests, and velocity offsets. Section~\ref{sec:observed_properties} presents the individual and stacked spectra, \lya--\heii kinematics, \lya EW and spatial constraints, and UV line ratios. Section~\ref{sec:contaminant_tests} evaluates possible contaminants and alternative ionizing sources. Section~\ref{sec:physical_interpretation} discusses the line ratios, and their connection to very low-metallicity or Pop~III-like ionizing sources. Section~\ref{sec:future_followup} outlines the spectroscopy needed to test the interpretation, and Section~\ref{sec:conclusions} summarizes the main results. Throughout this work, we adopt the Planck 2018 cosmology with $\Omega_{\rm m}=0.31$ and $H_{0}=67.7~{\rm km~s^{-1}~Mpc^{-1}}$ \citep{planck_collaboration_i_planck_2020}. All magnitudes are in the AB system \citep{Oke_and_Gunn_1983}. All number densities are comoving.

\section{Data and Parent Sample}
\label{sec:data}

We define a purity-oriented HETDEX LAE parent sample in the redshift interval where both \lya\ and \heii\ fall within our spectrograph's $3500$--$5500$~\text{\AA} wavelength window. This section summarizes the survey data, parent-sample cuts, and ancillary products used for validation. The candidate-selection procedure itself is described in Section~\ref{sec:candidate_selection}.

\subsection{HETDEX}
\label{subsec:hetdex_virus}

HETDEX \citep{gebhardt_hobby-eberly_2021} is an untargeted wide-field spectroscopic survey conducted with the upgraded Hobby--Eberly Telescope \citep[HET;][]{ramsey_early_1998,hill_hetdex_2021,hill_completion_2018}. Spectra are obtained with the Visible Integral-field Replicable Unit Spectrographs \citep[VIRUS;][]{hill_virus_2018,hill_hetdex_2021}, a highly replicated integral-field spectrograph that obtains spectra without imaging preselection. This untargeted spectroscopic design is central to the present search because it allows rare emission-line systems to be identified directly from their spectra.

VIRUS has 78 active integral-field units (IFUs) in the HET focal plane, corresponding to nearly 35,000 fibers. Each IFU covers $51\arcsec \times 51\arcsec$ on the sky with 448 fibers of diameter $1\farcs5$. At the sample median redshift, the $1\farcs5$  fiber diameter corresponds to approximately 12.8 physical kpc. The fiber fill factor in each IFU is $1/3$, so HETDEX uses a three-position dither pattern to fill the area of the IFUs. The spectra cover 3500--5500~\AA\ at resolving power $R\sim850$ \citep{hill_virus_2018}. At $1.9<z<2.3$, this wavelength range places \lya\ at 3525--4012~\AA\ and \heii\ at 4757--5413~\AA, placing both lines within the VIRUS spectral coverage. 

The HETDEX reduction and source-detection procedures are described by \citet{gebhardt_hobby-eberly_2021}. In brief, the pipeline calibrates the spectra, subtracts the sky, identifies emission-line and continuum sources, and provides catalog-level measurements and one-dimensional spectral extractions. The Emission Line eXplorer (ELiXer) software package classifies detected emission lines using spectroscopic information together with available imaging \citep{davis_hetdex_2023}. We use these catalog products to define the LAE parent sample and then search the extracted spectra for \heii emission.

\subsection{LAE Parent Sample}
\label{subsec:lae_parent_catalog}

We draw the parent sample from the HETDEX Data Release 5 (HDR5) source catalog \citep{mentuch_cooper_2026}. The goal is to define a high-confidence LAE sample for a rare-candidate search, not to construct a complete LAE luminosity function. This approach prioritizes purity, reliable spectral coverage, and uniform catalog quality.

We restrict the sample to sources with $1.9<z<2.3$, selecting high-confidence LAEs using ELiXer classification and the main HETDEX catalog quality flags \citep{mentuch_cooper_hetdex_2023,davis_hetdex_2023}. Specifically, we require the source to be classified as an LAE with $P_{\rm Ly\alpha}>0.8$.\footnote{Here $P_{\rm Ly\alpha}$ is ELiXer's empirical weighted-vote confidence score for the Ly$\alpha$ interpretation, combining the available spectroscopic and imaging diagnostics. We use $P_{\rm Ly\alpha}>0.8$ as a conservative, purity-oriented threshold. The threshold should not be interpreted as implying either an 80\% probability for each object or an 80\% pure parent sample.} Only detections satisfying the relevant HETDEX best-detection and quality-selection flags are retained in the sample. We also require Ly$\alpha$ to be detected with a signal-to-noise ratio ($\mathrm{S/N}>5$), an aperture correction factor \texttt{apcor} $>0.8$, a seeing/PSF FWHM $<1\farcs9$, and \texttt{linewidth} $<5.5$~\AA\null. The linewidth cut removes broad, poorly fit, or artifact-prone catalog detections and keeps the parent sample focused on narrow 
LAEs. The \texttt{apcor} cut removes detections requiring large aperture corrections, primarily sources falling near IFU edges where the flux recovery is less reliable. After these cuts, the parent sample contains 109,545 LAEs distributed across 4044 unique HETDEX observations.
 
\subsection{Spectral Products Used}
\label{subsec:spectral_products}

We use three HETDEX spectral products in our analysis: catalog line measurements, PSF-weighted one-dimensional spectral extractions, and fiber-level spectra. The catalog provides the source coordinates, shot ID, \lya\ wavelength, \lya\ flux, flux uncertainty, S/N, redshift, linewidth, aperture correction, and the observing-quality information used in the parent-sample selection. For each LAE, the catalog redshift predicts the approximate wavelength at which \heii\ would appear.

For the \heii\ search and subsequent line measurements, we use PSF-weighted, aperture-corrected one-dimensional spectra. The extraction combines the contributing fibers using the HETDEX PSF model, so the effective weight is dominated by fibers near the source position rather than by a uniform large aperture. These extractions provide the wavelength array, flux-density spectra, and uncertainty spectra used to identify candidate \heii emission and measure the stacked rest-frame UV line properties.

We also use fiber-level spectra during candidate validation to test whether the candidate \lya\ and \heii\ emission are spatially associated with the same source position rather than with a nearby object, a distant fiber, or localized artifact. The detailed validation checks are described in Section~\ref{sec:candidate_selection} and Appendix~\ref{app:fiber_checks}.

\subsection{Imaging and Ancillary Data}
\label{subsec:imaging_ancillary}

We use available imaging and ancillary data to characterize the candidates and assess possible contaminants. The optical imaging of the candidate objects comes from a variety of sources, including the Dark Energy Camera Legacy Survey \citep[DECaLS;][]{dey_2019}, the Hyper Suprime-Cam (HSC) Subaru Strategic Program \citep{aihara_2022}, and additional HSC images taken in support of HETDEX; these data are used to examine source morphology, obtain continuum-counterpart information, and identify nearby objects that could affect the HETDEX extraction. External spectroscopic, radio, and X-ray information provides additional tests for low-redshift interlopers and AGN contamination, as well as an independent context for the candidate validation. The contaminant tests are discussed in Section~\ref{sec:contaminant_tests}.

\section{Candidate Selection and He~\textsc{II} Search}
\label{sec:candidate_selection}

The candidate selection reduces the 109,545-object LAE parent sample to a small set of plausible \lya+ \heii\ systems with weak or absent UV metal-line emission. We scan the extracted spectra for emission near \heii, rank the strongest candidates, and apply visual, fiber-level, imaging, metal-line, AGN, and interloper checks. The selection prioritizes purity over completeness: it is designed to identify high-priority targets with strong candidate \heii, high \heii/Ly$\alpha$ line ratios, and weak metal lines, not to define a complete or unbiased population of all \heii emitters. Hybrid, self-enriched, embedded, lower-S/N, or metal-line-emitting Pop~III-active systems would generally fall outside this selection.

\begin{figure*}[t]
    \centering
    \includegraphics[width=\textwidth]{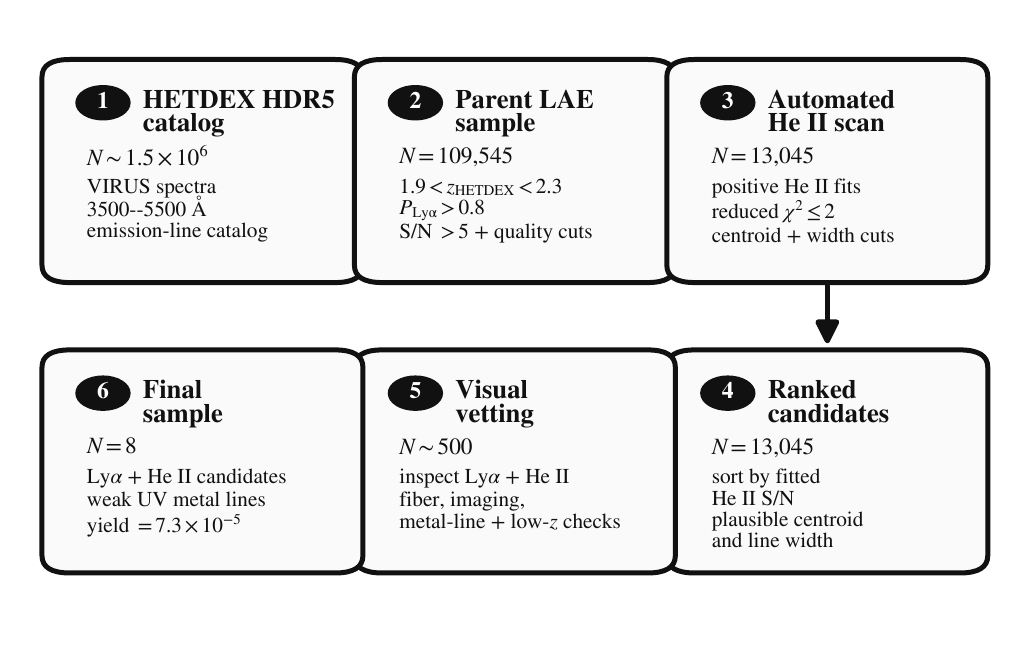}
    \caption{
    \textbf{Overview of the HETDEX Ly$\alpha$ + He~\textsc{II} candidate-selection pipeline.}
    We begin with the HETDEX HDR5 source catalog, define a high-confidence LAE parent sample at $1.9<z<2.3$, scan the extracted spectra for candidate He~\textsc{ii} $\lambda1640$ emission, rank candidates by fitted He~\textsc{ii} S/N, and visually vet the highest-ranked sources using spectral, fiber-level, imaging, metal-line, AGN, and low-redshift interloper checks. The final sample contains eight Ly$\alpha$ + He~\textsc{ii} candidates, corresponding to an observed yield of $7.3\times10^{-5}$ relative to the parent LAE sample. Since we target only the rarest He~\textsc{ii} + weak-metal-line objects, candidates with obvious strong UV metal lines are rejected from the sample.
    }
    \label{fig:sample_flowchart}
\end{figure*}

\subsection{{Search and Candidate Ranking}}
\label{subsec:search_motivation}

The search uses the LAE redshift to predict the expected observed wavelength of \heii. Because radiative transfer can shift Ly$\alpha$ from the systemic velocity, we allow the fitted \heii center to vary over 1630--1650~\AA\ in the rest frame. Figure~\ref{fig:sample_flowchart} summarizes the automated ranking and subsequent validation that lead to the final sample.
At $z\sim2$, Ly$\alpha$ velocity offsets relative to optically thin systemic tracers show substantial source-to-source variation. Typical offsets are $\sim 150-200~\kms$ for LAE samples and $\sim 400-450~\kms$ for more massive Lyman break galaxy (LBG) samples \citep{hashimoto_2015,steidel_2010}. The broader search window is therefore intended to retain objects with physically plausible Ly$\alpha$--systemic offsets rather than to assume that the Ly$\alpha$ catalog redshift is systemic.

We then examine the PSF-weighted one-dimensional spectrum around this wavelength window. The scan estimates the local continuum and noise from spectral windows on either side of the expected line position and measures the flux in the corresponding \heii\ window. This step provides an initial estimate of whether an emission feature is present near the expected \heii\ wavelength.

We also fit each \heii\ region with a Gaussian profile plus a local linear continuum. The fit returns the line center, observed-frame width, integrated flux, flux uncertainty, S/N, and reduced $\chi^{2}$. We keep candidates with successful positive-flux fits, acceptable fit quality, fitted centers close to the expected \heii\ wavelength, and widths consistent with resolved or marginally resolved emission while rejecting both single-pixel artifacts and very broad features. Specifically, we require reduced $\chi^{2}\leq2$, an observed line width between 1.7 and 7~\AA, and a fitted line center within $\sim10$~\AA\ of the expected \heii\ position in the observed frame; at $z =2$, this window corresponds to $\sim600~\kms$. The centroid window allows for velocity offsets between resonant \lya and non-resonant emission while rejecting features far from the predicted \heii wavelength. These automated requirements produce an initial list of approximately 13,000 spectra with positive, acceptable \heii fits. We rank this initial list by fitted \heii S/N and carry forward the highest-ranked $\sim$500 objects for detailed inspection. The inspected ranking spans fitted \heii\ S/N $=7.50$ at rank 1 to $\simeq1.85$ at rank 500. The cutoff at 500 is used as a conservative inspection limit rather than as a completeness threshold; below this rank, the \heii feature is very weak, and the number of noise peaks, artifacts, and single-fiber features increases rapidly.

\subsection{Visual Vetting}
\label{subsec:visual_vetting}

We visually inspect the $\sim$500 highest-ranked objects to determine whether the candidate \heii\ feature is credible and physically associated with its LAE\null. For each candidate, we check whether the emission lies near the expected \heii\ wavelength, whether it extends over more than a single noisy pixel, whether the \lya\ line and adopted redshift provide a consistent two-line interpretation, and whether the spectrum shows obvious UV metal-line or AGN-like features inconsistent with the weak-metal-line class targeted here.

We reject artifacts, poor extractions, spatially inconsistent line emission, obvious foreground interpretations, and spectra with clear \nv\ $\lambda1240$, \civ\ $\lambda1549$, or \oiii\ $\lambda1663$ emission. These are the three primary rest-UV metal-line diagnostics used for selection and the measurement. We require the retained candidates to be consistent with a common \lya and \heii source position at the scale of the VIRUS fibers. The detailed imaging and interloper checks are in Appendix~\ref{app:ancillary_interlopers}.

\subsection{Final Candidate Sample}
\label{subsec:final_candidate_sample}

After automated ranking and visual vetting, including rejection of obvious metal-line-rich or AGN-like spectra, we identify eight \lya+ \heii\ candidates with no clear UV metal-line emission. Most of these candidates are drawn from the top 50 objects in the automated \heii S/N ranking, but the inspection was extended to the top 500 to check for credible lower-ranked sources. These objects form the final sample analyzed in the remainder of the paper. Table~\ref{tab:final_candidates} lists their HETDEX catalog identifiers, coordinates, \lya-based redshifts, and \lya luminosities.

Representative rejected cases like a noise spike, a double-LAE detection, a low-redshift interloper, and a metal-line-rich source are shown in Appendix~\ref{app:rejected_examples}.

\begin{deluxetable}{crrrr}
\tablecaption{Final Ly$\alpha$ + He~\textsc{ii} Candidate Sample}
\label{tab:final_candidates}
\tablehead{
\colhead{detectid} &
\colhead{$\alpha_{\rm J2000}$ (deg)} &
\colhead{$\delta_{\rm J2000}$ (deg)} &
\colhead{$z$} &
\colhead{\begin{tabular}{c}
$\log L_{\mathrm{Ly}\alpha}$ \\
$(\mathrm{erg~s^{-1}})$
\end{tabular}}
}
\startdata
3005020033 & 165.19334 &  50.792854 & 2.213 & 42.93 \\
3007998098 & 221.16264 &  53.845905 & 1.994 & 43.32 \\
3009751099 &  19.451492 &   0.471328 & 2.038 & 42.92 \\
3010547400 &  20.290672 &  -1.048341 & 2.053 & 42.94 \\
4019081078 & 233.91087 &  51.616085 & 1.994 & 42.99 \\
4028177445 & 228.20595 &  50.607033 & 2.184 & 42.76 \\
4028632963 & 211.51462 &  52.452236 & 2.106 & 43.20 \\
5001528748 & 170.90740 &  49.509910 & 2.204 & 42.65 \\
\enddata
\tablecomments{Redshifts are Ly$\alpha$-based HETDEX catalog redshifts. The catalog \lya line S/N values for the candidates range from $5.3$--12.2. The calculated \heii line S/N values for the candidates range from $2.4$--3.6. The median Ly$\alpha$ luminosity of the sample is $\log L_{\mathrm{Ly}\alpha}=42.94$.} 
\end{deluxetable}

\subsection{Selection Function and Rarity}
\label{subsec:selection_rarity}

The final sample is intentionally conservative and biased toward visually credible, high-priority \heii\ candidates without obvious UV metal-line emission. Because the selection combines automated thresholds, S/N ranking, and visual vetting, we do not derive a completeness-corrected \heii\ luminosity function, but instead provide a first-order estimate of the candidate comoving number density for comparison with theory and previous searches.

We estimate the parent LAE number-density normalization from the HETDEX Ly$\alpha$ luminosity function of \citet{zhang_2021}. Their faint-component Schechter fit gives $\log(L^\ast_{\rm Ly\alpha}/{\rm erg~s^{-1}})\simeq42.87$, and $\log(\phi^\ast/{\rm Mpc}^{-3})\simeq-3.41$, corresponding to $\phi^\ast\simeq3.9\times10^{-4}~{\rm Mpc}^{-3}$. Adopting this value as an approximate LAE number-density normalization and scaling by $8/109{,}545$ gives $n_{\rm cand}\simeq2.9\times10^{-8}~{\rm Mpc}^{-3}$, or $\sim30~{\rm Gpc}^{-3}$.

This estimate should be interpreted cautiously. False positives or misidentifications would lower the true density of systems matching this \heii$+$ weak-metal-line selection, whereas incompleteness from the exclusion of hybrid, self-enriched, lower-S/N, metal-line-emitting, embedded, or non-LAE systems would raise the density of the broader physical population. Because neither correction is quantified, their net effect is unknown. We therefore treat $n_{\rm cand}\sim30~{\rm Gpc}^{-3}$ as a first-order normalization for the observed candidate class, not as a strict lower limit or a completeness-corrected occurrence rate. A more rigorous estimate requires completeness modeling, which we defer to future work.

\section{Spectral Extraction, Stacking, and Line Measurements}
\label{sec:extraction_stacking_measurements}

The quantitative measurements in this paper are based on the stacked spectra of the eight candidates. Individual spectra are used for candidate selection and validation, but their continuum levels are generally too weak for robust object-by-object equivalent-width (EW) measurements. We therefore use the individual spectra to define and check the sample, and we use the eight-object stack for the line luminosities, EWs, line ratios, velocity widths, and metal-line upper limits reported below.

\subsection{{Spectral Extraction and Stacking}}
\label{subsec:psf_weighted_extraction}

\begin{figure*}[t]
    \centering \includegraphics[width=\textwidth]{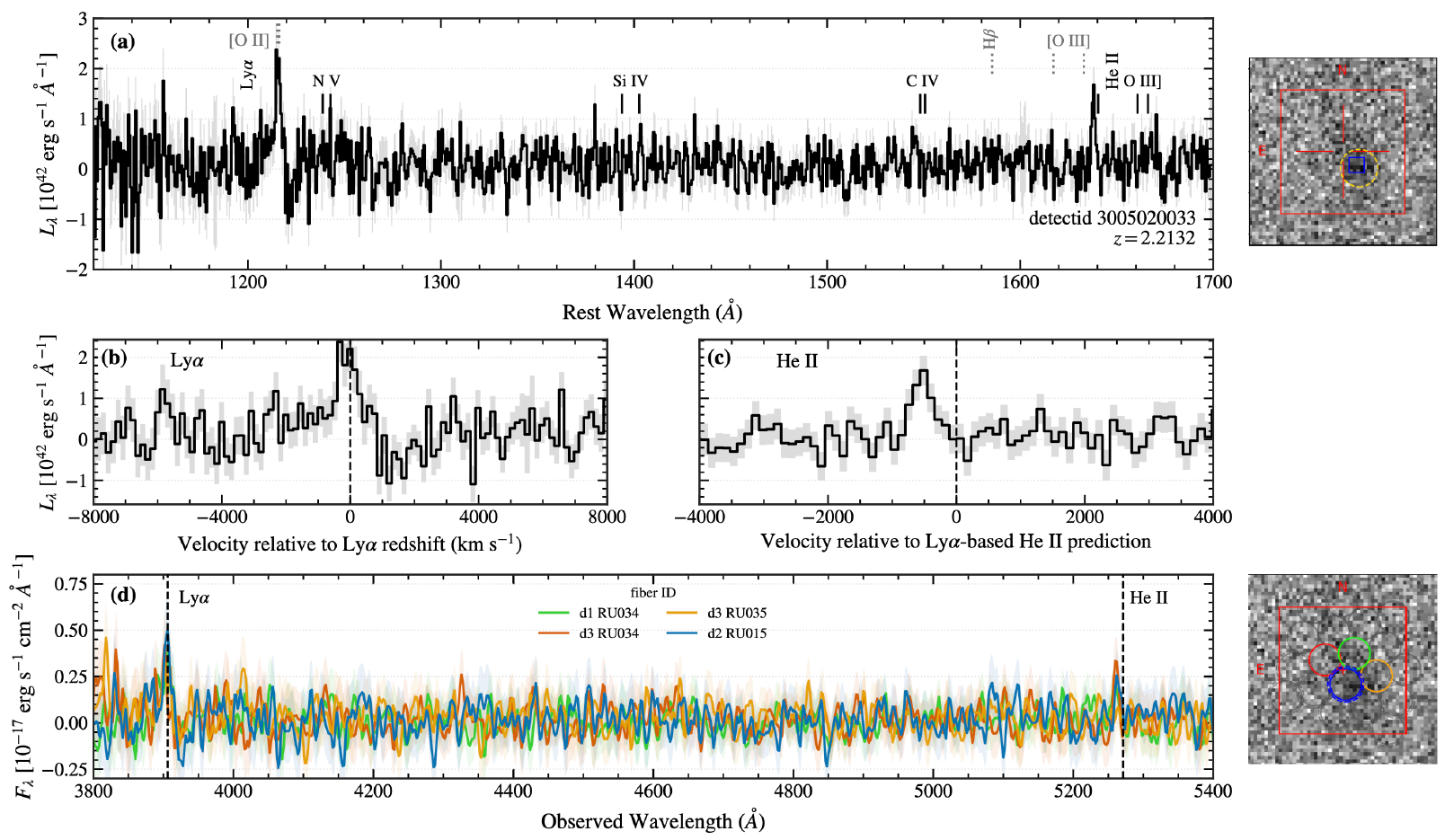}
    \caption{
    \textbf{Example Ly$\alpha$ + \ion{He}{2} candidate}, detectid 3005020033.
    Panel (a): the PSF-weighted, aperture-corrected rest-frame UV spectrum. Solid black vertical markers indicate major rest-frame UV transitions assuming the Ly$\alpha$ interpretation, while dotted gray markers show where the strongest rest-frame optical lines would appear under the alternative low-redshift [O~\textsc{ii}] interloper hypothesis. The inset on the right displays the HSC $r$-band (mag $\sim 25$) $8\arcsec \times 8\arcsec$ cutout. The gold ellipse marks the source extractor continuum detection, and the small blue square marks the corresponding HSC $r$-band catalog detection center.
    Panels (b) and (c): the Ly$\alpha$ and \ion{He}{2} regions in velocity space, with $v=0$ corresponding to the wavelength predicted from the HETDEX catalog's Ly$\alpha$-based redshift.
    The measured Ly$\alpha$--He~\textsc{ii} offset is $\Delta v_{\rm Ly\alpha-HeII}=+404^{+86}_{-95}~\mathrm{km~s^{-1}}$. The immediate blueward excess is localized rather than a broad baseline or common sky residual and may reflect source-associated profile structure or complex kinematics (see Appendix~\ref{app:figure2_blue}). Panel (d): the nearby individual fiber spectra and positions, which are spatially consistent with a common Ly$\alpha$ and candidate \ion{He}{2} source (see Appendix~\ref{app:fiber_checks}). The fiber spectra are smoothed with a one-pixel Gaussian kernel. The right inset displays the nearby fiber positions, plotted to the correct scale in circles with colors corresponding to the spectra in panel (d). The typical astrometric uncertainty in HETDEX is $\sim 0\farcs7$ for low-S/N detections.
    }
    \label{fig:example_candidate_3005020033}
\end{figure*}

For each candidate, we use the PSF-weighted spectrum extracted at the LAE position. Figure~\ref{fig:example_candidate_3005020033} illustrates the validation for one representative source: the one-dimensional spectrum and velocity panels locate \lya and candidate \heii, while the imaging, fiber map, and individual fiber spectra test whether both lines are associated with the same HETDEX source. The complete fiber-level and extraction-center checks are presented in Appendix~\ref{app:fiber_checks}. The corresponding spectra for the remaining candidates are shown in Appendix~\ref{app:remaining_candidate_checks}.

We convert the observed flux-density spectra to rest-frame luminosity density before stacking. We then shift each spectrum to its rest frame and interpolate it onto a common wavelength grid.

We construct two rest-frame versions of the spectra. For measurements centered on \heii, we use preliminary fitted \heii centroids to build \heii-aligned spectra, placing the \heii emission at its rest wavelength before stacking. This alignment reduces line broadening caused by velocity offsets between the apparent resonant \lya line and the fitted \heii centroid. It also places the nearby rest-frame UV metal-line regions in the same velocity frame used for the \heii measurement. The \heii-aligned stack is therefore used for the \heii profile, UV metal-line limits, and the systemic-frame \lya--\heii velocity-offset measurement. For measurements centered on \lya, we instead use the HETDEX catalog \lya-based redshift and measure \lya in the \lya-aligned stack; this version preserves the \lya profile and flux for the fiducial luminosity and EW measurements.

We stack the eight final candidates listed in Table~\ref{tab:final_candidates} in luminosity density using the Tukey biweight estimator \citep{beers_1990}, without continuum normalization; this choice avoids introducing additional uncertainties associated with the relative faintness of the continua. All eight spectra contribute at the wavelengths of \lya, \nv, \civ, \heii, and \oiii.

We do not report individual-object EWs or line ratios because the continua have low S/N; the stack provides the quantitative line and continuum measurements. Since the candidates are selected partly on emission-like features near the expected \heii wavelength, the stack is not treated as an independent measurement of the false-positive rate; we address this point with null tests in Section~\ref{subsec:noise_skyline}.

\subsection{Continuum Estimation and Line Fitting}
\label{subsec:line_fitting}

We adopt Markov Chain Monte Carlo (MCMC) sampling \citep{foremanmackey_2013} with a Gaussian line-profile model as the fiducial measurement method for the stacked \heii and \lya lines. Fixed-window non-parametric integration is used as a robustness check, not as the primary measurement.

The continuum is estimated locally around each line using a linear model constrained by sidebands. For \heii, the fit is performed over 1610--1670~\AA, with the line center allowed to vary over 1629--1651~\AA\null. The \heii EW is reported over the same 1629--1651~\AA\ line window. For \lya, the fit is performed over 1195--1235~\AA, with the line center allowed to vary over 1208--1224~\AA. The \lya EW is reported over 1205--1227~\AA.

We do not reconstruct flux absorbed blueward of \lya; the reported \lya luminosity and EW are observed quantities. We test red side only continua, split-Gaussian and red wing only Gaussian profiles, and non-parametric integrations, as described in Appendix~\ref{app:lya_profile_robustness}.

For each line, the EW is defined as positive for emission and is computed as the line luminosity divided by the fitted local continuum. The fitted \heii FWHM is measured as an observed stack width in velocity units. We also report an intrinsic FWHM after correcting for the representative VIRUS instrumental resolution, as described in Section~\ref{subsec:stacked_spectrum_all8}. The \heii/\lya ratio is computed from line luminosities, not from EWs: \heii is measured in the \heii-aligned stack, while \lya is measured in the \lya-aligned stack for the same eight objects.

For the velocity-offset measurement, we refit the \lya and \heii centroids in each PSF-weighted spectrum. We define
\begin{equation}
\Delta v_{\rm Ly\alpha-HeII}=c\,\frac{z_{\rm Ly\alpha}-z_{\rm HeII}}{1+z_{\rm HeII}},
\end{equation}
so that positive values indicate \lya redshifted relative to \heii. The \heii centroid is treated as an approximate systemic reference. The centroid-fitting and posterior propagation are detailed in Appendix~\ref{app:lya_kinematics}.

\subsection{Uncertainties and Robustness Tests}
\label{subsec:measurement_uncertainties}

We combine MCMC posterior uncertainties with object-level bootstrap and leave-one-out resampling, and estimate method-dependent systematics from the scatter across stacking, continuum, profile, and integration choices. The \heii measurement is stable across this grid. The \lya luminosity is substantially more robust than its EW, which remains continuum- and profile-limited. Full resampling and robustness results are given in Appendices~\ref{app:stack_robustness} and \ref{app:lya_profile_robustness}.

\subsection{Metal-line Upper Limits}
\label{subsec:metal_upper_limits}

We use the \heii-aligned stack to place upper limits on nearby rest-frame UV metal lines. The limits reported in this paper are for \nv\ $\lambda1240$, \civ\ $\lambda1549$, and \oiii\ $\lambda1663$. These transitions are important because the existence of strong high-ionization metal emission would favor an interpretation involving AGN, shocks, or metal-enriched stellar populations over a very low-metallicity interpretation. The lines also provide a stack-level check on the visual selection, which intentionally excludes candidates with obvious individual metal-line counterparts.

For each metal line, we estimate a local continuum from sidebands around the expected rest wavelength and evaluate a Gaussian line model with the line center fixed to the appropriate wavelength. Because these metal lines are not significantly detected, we do not interpret positive noise fluctuations as line detections. Instead, we assume a Gaussian velocity width tied to the measured observed \heii stack width and compute the $3\sigma$ luminosity upper limit from the local uncertainty over that line profile.

We divide each metal-line luminosity limit by the measured \heii luminosity from the same \heii-aligned stack to obtain upper limits on \nv/\heii, \civ/\heii, and \oiii/\heii. We emphasize that these values are upper limits, not detections. Any uncertainty associated with the numerical limit reflects the uncertainty in the limit calculation and the \heii normalization, not a measured metal-line luminosity.

\section{Observed Properties}
\label{sec:observed_properties}

We now present the \heii spectral region and \lya--\heii velocity offsets for the individual candidates, followed by the final stacked spectrum and the additional \lya profile and spatial measurements. The individual spectra show why the objects are promising targets for future observations and why stacking is required for the line-ratio and EW measurements.

\subsection{Individual Candidate Spectra}
\label{subsec:individual_candidates}

\begin{figure}[t]
    \centering
    \includegraphics[width=\columnwidth]{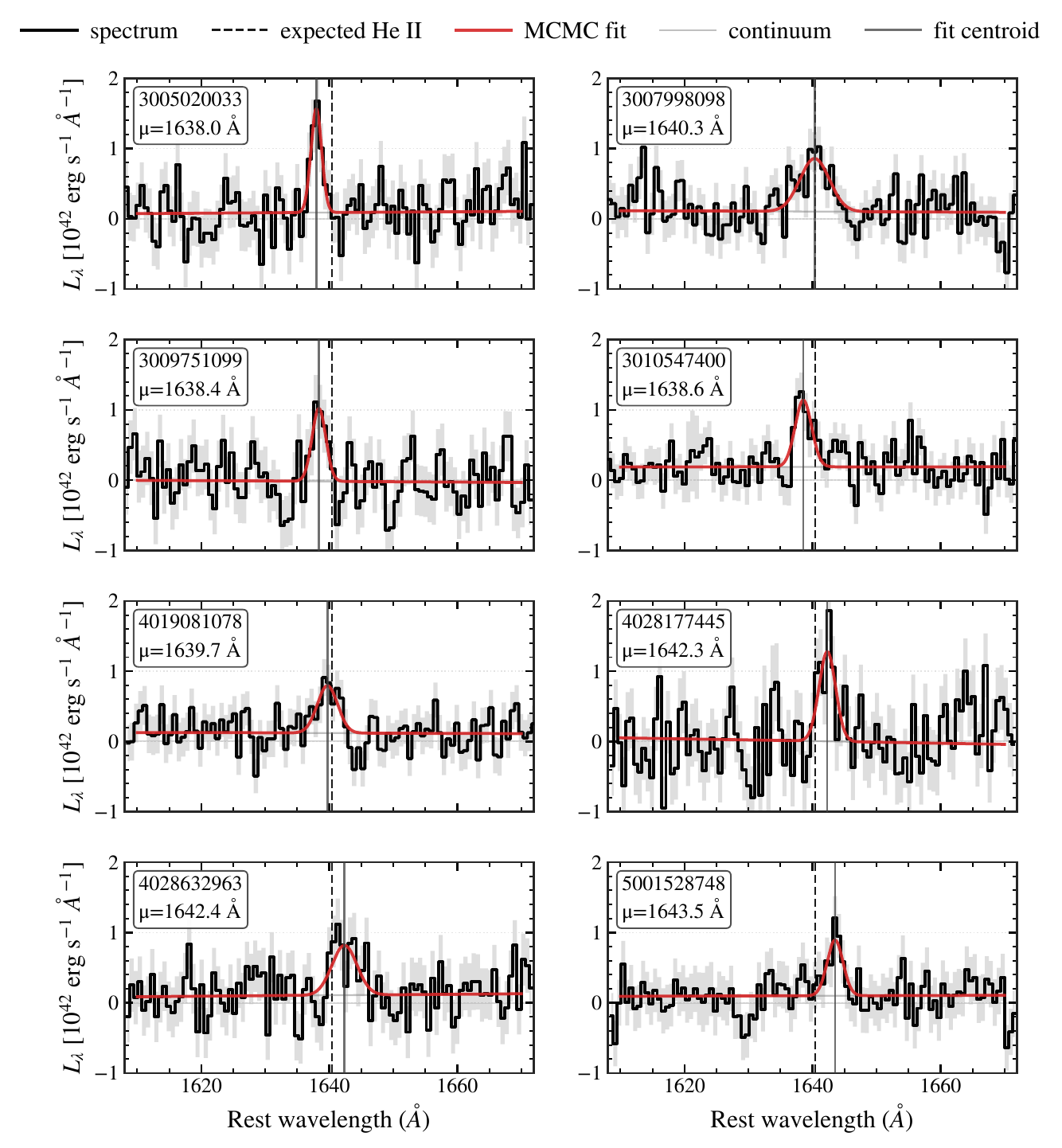}
    \caption{
    \textbf{\ion{He}{2} $\lambda1640$ spectral region for the eight final Ly$\alpha$ + \ion{He}{2} candidates.}
    The spectra are displayed in the rest frame defined by the catalog Ly$\alpha$ redshift.
    The PSF-weighted, aperture-corrected extracted spectra are in black, and gray bands show the $1\sigma$ uncertainties.
    The dashed vertical line marks the expected \ion{He}{2} position from the Ly$\alpha$-based redshift, while the solid vertical line shows the fitted \ion{He}{2} centroid used for rest-frame alignment when available. The corresponding Ly$\alpha$--He~\textsc{ii} offsets and centroid uncertainties are listed in Section~\ref{subsec:individual_candidates}.
    The individual spectra are used for candidate validation and stack construction, not for EW or line-ratio measurements.
    }
    \label{fig:all_eight_heii}
\end{figure}

The eight candidates span a range of line strengths and data quality, so we first examine their individual \heii regions and \lya--\heii kinematics before turning to the stack. Figure~\ref{fig:all_eight_heii} shows the \heii spectral region for the final eight candidates. The spectra are presented in the rest frame defined by the catalog \lya redshift. The expected \heii wavelength from the \lya-based redshift is marked in each panel, and the fitted \heii centroid used for alignment is shown. Offsets between these two markers are not by themselves evidence against the \lya+ \heii interpretation, because \lya is resonant and can be shifted relative to \heii emission.

Using the sign convention in Section~\ref{subsec:line_fitting}, the measured offsets are $+404^{+86}_{-95}$, $+35^{+108}_{-100}$, $+428^{+188}_{-138}$, $+347^{+86}_{-94}$, $+157^{+103}_{-98}$, $-381^{+116}_{-117}$, $-352^{+145}_{-144}$, and $-526^{+132}_{-110}~\kms$ for detectids 3005020033, 3007998098, 3009751099, 3010547400, 4019081078, 4028177445, 4028632963, and 5001528748, respectively. The $95\%$ interval excludes zero for the three most positive and all three negative cases; detectids 3007998098 and 4019081078 remain consistent with zero.

\subsection{Final Eight-object Stack}
\label{subsec:stacked_spectrum_all8}

\begin{figure*}[t]
    \centering
    \includegraphics[width=\textwidth]{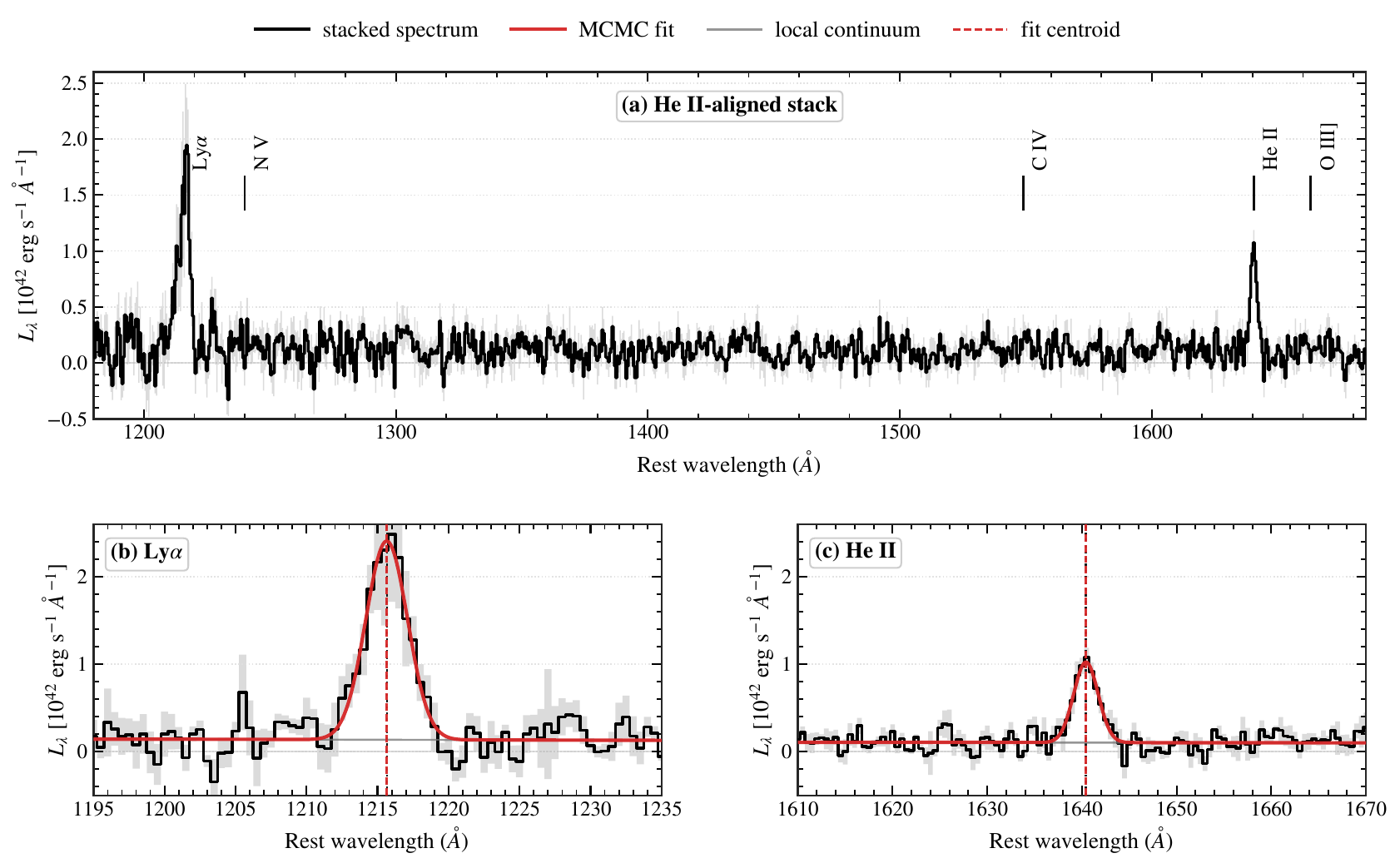}
    \caption{
    \textbf{Final stacked spectra.}
    Panel (a): the He~\textsc{ii}-aligned rest-frame UV stack, with major rest-frame UV transitions marked.
    Panel (b): the Ly$\alpha$ region measured from the separate Ly$\alpha$-aligned stack.
    Panel (c): an expanded view of the He~\textsc{ii} $\lambda1640$ region from panel (a).
    Gray bands show the stack uncertainties, and red curves indicate the fiducial Gaussian MCMC fits.
    The He~\textsc{ii}-aligned stack is used to measure the He~\textsc{ii} profile and the N~\textsc{v}, C~\textsc{iv}, and O~\textsc{iii}] upper limits; the Ly$\alpha$-aligned stack is used for the Ly$\alpha$ luminosity and EW. The stack shows clear He~\textsc{ii} emission and no comparably strong N~\textsc{v}, C~\textsc{iv}, or O~\textsc{iii}] emission. Ly$\alpha$ measured separately in the He~\textsc{ii}-aligned stack gives $\Delta v_{\rm Ly\alpha-HeII}=+13^{+264}_{-173}~\mathrm{km~s^{-1}}$, consistent with zero.
    }
    \label{fig:heii_aligned_stack}
\end{figure*}

The final eight-object stack reduces the impact of noise and continuum uncertainty in the individual spectra. Figure~\ref{fig:heii_aligned_stack} shows the stacked rest-frame UV spectrum and the line-region fits used for the main measurements. The \heii-aligned stack is used for the \heii profile and nearby metal-line limits, while the \lya-aligned stack is used for the \lya luminosity entering the \heii/\lya ratio.

The stack shows a clear \heii feature and no significant \nv, \civ, or \oiii emission. Using the biweight stack, wide continuum window, and Gaussian MCMC as the fiducial measurement, we find ${\rm EW}_{\rm HeII,rest}=28.7\pm5.6_{\rm stat}\pm3.7_{\rm sys}~{\rm \AA}$, $L_{\rm HeII}=(2.98\pm0.32_{\rm stat}\pm0.21_{\rm sys})\times10^{42}~{\rm erg~s^{-1}}$, and an observed \heii width of ${\rm FWHM}_{\rm obs}=559\pm84_{\rm stat}\pm7_{\rm sys}~\kms$. Assuming the VIRUS red-side resolving power of $R=900$ \citep{hill_hetdex_2021} corresponds to an instrumental FWHM of $\simeq333~\kms$, the intrinsic width of the \heii line is $449\pm105_{\rm stat}\pm9_{\rm sys}~\kms$. This width is far below the broad-line regime typical of unobscured broad-line AGN, but significantly broader than a single-pixel noise spike. 

The \lya-aligned stack gives ${\rm EW}_{\rm Ly\alpha,rest}=61.8\pm14.7_{\rm stat}\pm57.4_{\rm sys}~{\rm \AA}$, $L_{\rm Ly\alpha}=(8.37\pm2.06_{\rm stat}\pm1.35_{\rm sys})\times10^{42}~{\rm erg~s^{-1}}$, and a resulting line-ratio of ${\rm HeII/Ly\alpha}=0.354\pm0.079_{\rm stat}\pm0.051_{\rm sys}$. The large systematic uncertainty on the \lya EW reflects its greater sensitivity to continuum placement and alignment choice. Numerically, the combined uncertainty is approximately $96\%$ of the Ly$\alpha$ EW but only approximately $29\%$ of the Ly$\alpha$ luminosity; the luminosity is therefore the substantially more reliable Ly$\alpha$ measurement. The full continuum/profile grid is given in Appendix~\ref{app:lya_profile_robustness}. The \heii measurement and \heii/\lya ratio are more stable across the robustness tests (see Appendix~\ref{app:stack_robustness}). In the \heii-aligned stack, the \lya centroid gives $\Delta v_{\rm Ly\alpha-HeII}=+13^{+264}_{-173}~\kms$. The offset from the stack is therefore consistent with zero even though the individual objects span a broad range of positive and negative values.

The metal-line constraints provide the main additional diagnostic. Using the observed \heii stack width to set the assumed line-width scale, we obtain $3\sigma$ upper limits of \nv/\mheii\ $<0.167$, \civ/\mheii\ $<0.128$, and \oiii/\mheii\ $<0.140$. Table~\ref{tab:heii_measurements} summarizes the final stacked measurements and upper limits. The matched-filter significances of these metal-line regions are below the detection threshold, so all three are treated as non-detections. The absence of comparably strong \nv, \civ, and \oiii emission is central to the contaminant tests in Section~\ref{sec:contaminant_tests}.

\begin{deluxetable}{ll}
\tablecaption{Final Eight-object Stack Measurements}
\label{tab:heii_measurements}
\tabletypesize{\scriptsize}
\setlength{\tabcolsep}{2pt}
\tablehead{
\colhead{Quantity} &
\colhead{Value}
}
\startdata
He~\textsc{ii} $\lambda1640$ EW$_{\mathrm{rest}}$ & $28.7\pm6.7~\mathrm{\mathring{A}}$ $(5.6_{\mathrm{stat}},\,3.7_{\mathrm{sys}})$ \\
$L_{\mathrm{HeII}}$ & $(2.98\pm0.38)\times10^{42}~\mathrm{erg~s^{-1}}$ $(0.32_{\mathrm{stat}},\,0.21_{\mathrm{sys}})$ \\
Ly$\alpha$ EW$_{\mathrm{rest}}$ & $61.8\pm59.3~\mathrm{\mathring{A}}$ $(14.7_{\mathrm{stat}},\,57.4_{\mathrm{sys}})$ \\
$L_{\mathrm{Ly}\alpha}$ & $(8.37\pm2.46)\times10^{42}~\mathrm{erg~s^{-1}}$ $(2.06_{\mathrm{stat}},\,1.35_{\mathrm{sys}})$ \\
He~\textsc{ii}/Ly$\alpha$ & $0.354\pm0.094$ $(0.079_{\mathrm{stat}},\,0.051_{\mathrm{sys}})$ \\
$\Delta v_{\rm Ly\alpha-HeII}$ (He~\textsc{ii}-aligned stack) & $+13^{+264}_{-173}~\mathrm{km~s^{-1}}$ \\
He~\textsc{ii} $\lambda1640$ FWHM$_{\mathrm{obs}}$ & $559\pm84~\mathrm{km~s^{-1}}$ $(84_{\mathrm{stat}},\,7_{\mathrm{sys}})$ \\
He~\textsc{ii} $\lambda1640$ FWHM$_{\mathrm{int}}$ ($R=900$) & $449\pm105~\mathrm{km~s^{-1}}$ $(105_{\mathrm{stat}},\,9_{\mathrm{sys}})$ \\
$L_{\mathrm{N\,V},3\sigma}$ & $<4.97\times10^{41}~\mathrm{erg~s^{-1}}$ \\
$L_{\mathrm{C\,IV},3\sigma}$ & $<3.82\times10^{41}~\mathrm{erg~s^{-1}}$ \\
$L_{\mathrm{O\,III]},3\sigma}$ & $<4.15\times10^{41}~\mathrm{erg~s^{-1}}$ \\
N~\textsc{v}/He~\textsc{ii} & $<0.167$ \\
C~\textsc{iv}/He~\textsc{ii} & $<0.128$ \\
O~\textsc{iii}]/He~\textsc{ii} & $<0.140$ \\
\enddata
\tablecomments{
Fiducial measurements use the biweight stack, wide local continuum windows, and Gaussian MCMC fitting. He~\textsc{ii} $\lambda1640$ and metal-line limits are measured from the He~\textsc{ii} $\lambda1640$-aligned stack; the fiducial Ly$\alpha$ luminosity and EW are measured from the Ly$\alpha$-aligned stack, while the velocity offset is measured from Ly$\alpha$ in the He~\textsc{ii}-aligned stack. Metal-line luminosities and ratios are $3\sigma$ upper limits. The luminosity limits are integrated over a Gaussian profile with width fixed to the measured observed He~\textsc{ii} $\lambda1640$ stack width. The intrinsic He~\textsc{ii} $\lambda1640$ FWHM is an approximate correction to the observed stacked width with $R=900$. A quantitative metallicity measurement [$12+\log(\mathrm{O/H})$] requires rest-optical spectroscopy to constrain both gas-phase metal abundances and the ionization parameter.
}
\end{deluxetable}

The \lya spatial analysis classifies one of the eight candidates, detectid 3007998098, as resolved under the adopted criteria of \citet{mentuch_cooper_2026_LAN}. This source has an exponential scale length of $r_s=17.20\pm0.25$~kpc and an isophotal radius of $r_{\rm iso}=41.84$~kpc. The remaining seven candidates do not satisfy the resolved-source criterion. The individual \heii features are too weak for a corresponding quantitative size measurement. The candidate-level statistics are presented in Appendix~\ref{app:lya_spatial}.

\subsection{Equivalent-width and Line-ratio Diagnostics}
\label{subsec:ew_line_ratios}

\begin{figure*}[t]
\centering
\includegraphics[width=0.8\textwidth]{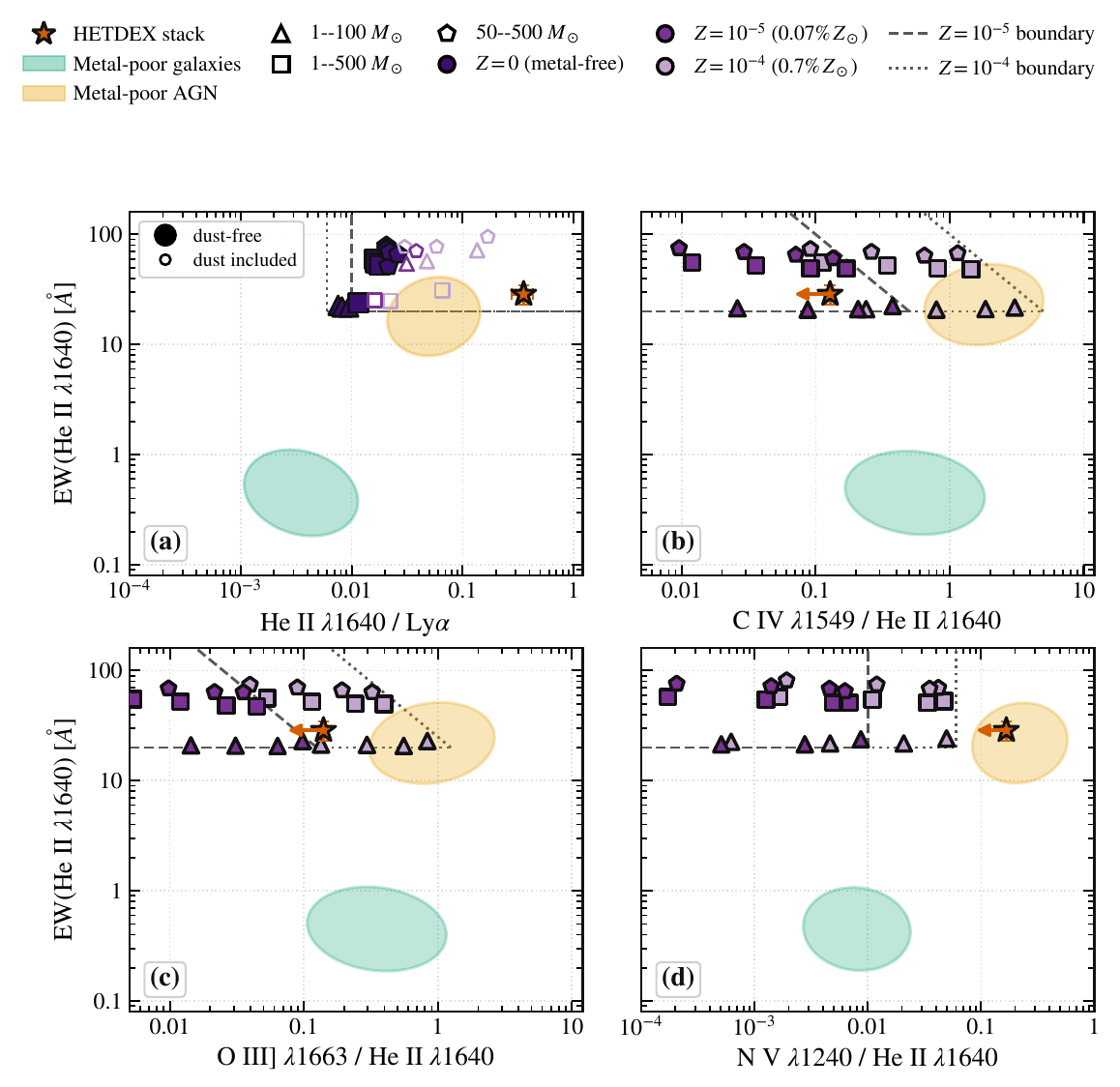}
\caption{
\textbf{Rest-frame UV diagnostic comparison for the final HETDEX Ly$\alpha$ + \ion{He}{2} stack and the \texttt{Cloudy} models of \citet{nakajima_maiolino_2022}.}
Panel (a): \heii EW versus He~\textsc{ii}/Ly$\alpha$. Larger symbols represent dust-free models; smaller hollow symbols show results that include dust.
Panels (b)--(d): the corresponding metal-line upper-limit diagnostics for \civ/\mheii, \oiii/\mheii, and \nv/\mheii.
The orange star marks the final stack.
Leftward arrows in panels (b)--(d) indicate $3\sigma$ upper limits on the metal-line ratios.
Purple model points show the Pop~III/very metal-poor models, with color indicating metallicity: dark purple for $Z=0$, purple for $Z=10^{-5}$, and light purple for $Z=10^{-4}$; marker shape indicates the model mass scale.
Gray line boundaries indicate the approximate Pop~III/very metal-poor diagnostic regions, with dashed lines for $Z=10^{-5}$ and dotted lines for $Z=10^{-4}$.
Schematic ovals in green and yellow indicate the approximate regions of low-metallicity galaxies and AGN-like sources, based on the comparison models in \citet{nakajima_maiolino_2022}; these regions are intended only as visual context, not as formal classification boundaries.
The final stack occupies a region with large He~\textsc{ii} EW, high He~\textsc{ii}/Ly$\alpha$, and weak metal-line emission. {Additional Ly$\alpha$ suppression would move the model points rightward in panel~(a). See Section~\ref{subsec:cloudy_grids} for further discussion.}
}
\label{fig:heii_diagnostic_grid}
\end{figure*}

Figure~\ref{fig:heii_diagnostic_grid} places the final stack as an orange star in the diagnostic space that motivated the final visual selection: \heii EW, \heii/\lya, and metal-line strength relative to \heii. The comparison \texttt{Cloudy} models from \citet{nakajima_maiolino_2022} show why this combination of axes is useful: strong \heii and high \heii/\lya trace a hard ionizing spectrum, while low \civ/\heii, \oiii/\heii, and \nv/\heii ratios test whether the hard spectrum is accompanied by very metal-poor gas. The purple model points indicate the Pop~III/very metal-poor models, while the gray dashed and dotted curves show the corresponding approximate low-metallicity diagnostic boundaries for $Z=10^{-5}$ and $Z=10^{-4}$, respectively. Schematic ovals in green and yellow indicate the approximate regions of low-metallicity galaxies and AGN-like sources and are intended only as visual context, not as formal boundaries.

Panel (a) shows that the final stack has both large \heii EW and high \heii/\lya. The high ratio is driven in part by intrinsically strong \heii. Additional Ly$\alpha$ suppression by dust or redistribution by radiative transfer effects remains plausible (see Section \ref{subsec:cloudy_grids}).
Panels (b)--(d) add the key constraint: the stack does not show correspondingly strong \civ, \oiii, or \nv emission. This aspect matters because many AGN, shock, and metal-enriched stellar scenarios can produce \heii, but they often also produce detectable UV metal lines with line widths comparable to or broader than the intrinsic \heii\ FWHM\null. The final stack occupies the part of the diagnostic space associated with very hard ionizing spectra in low-metallicity gas. The comparison to photoionization models does not establish a Pop~III origin, but it does identify the stack as a high-priority target for the contaminant tests and future observations, as discussed below. 

\section{Contaminant Tests}
\label{sec:contaminant_tests}

The final stacked spectrum requires an explanation for \heii emission that is strong relative to \lya and not accompanied by significant \nv, \civ, or \oiii emission. The final visual selection contributes to this line pattern by rejecting candidates with obvious strong UV metal lines or AGN-like spectra before stacking. Because \heii can arise from several astrophysical and observational sources, we examine whether the stack could be dominated by noise, low-redshift interlopers, AGN, fast shocks, cooling radiation, or Wolf--Rayet and other metal-enriched stellar populations. These tests do not assign a unique physical origin to every candidate, but test whether standard contaminants naturally reproduce the stacked line pattern.

\subsection{Noise, Artifacts, and Skyline Residuals}
\label{subsec:noise_skyline}

Noise fluctuations, skyline residuals, bad pixels, and problematic flatfielding can mimic weak emission features in individual spectra. This failure mode is especially important because the automated scan for Pop~III candidates intentionally searches for positive features near the expected \heii wavelength. We therefore do not treat the automated scan, by itself, as proof that a feature is astrophysical; it is only a starting point.

The final candidates pass fiber-level spatial checks, the stack remains stable when individual objects are removed, and blank-window tests on the same eight spectra do not reproduce a comparable positive feature. Parent-sample null searches also show that selecting and centroid-aligning the strongest noise features can create an artificial peak; consequently, we do not use the aligned stack alone to infer a false-positive rate. The complete validation and null-test procedures are given in Appendices~\ref{app:fiber_checks} and \ref{app:null_tests}. Deeper spectroscopy remains necessary to confirm the individual \heii features.

\subsection{Low- and High-redshift Interlopers}
\label{subsec:lowz_interlopers}

A low-redshift emission-line galaxy can mimic a \lya+ \heii source if [O~\textsc{ii}] $\lambda3727$ is misidentified as \lya \citep[e.g.,][]{leung_2017}. This failure mode is relevant for Ly$\alpha$+\heii searches because, under the [O~\textsc{ii}] interpretation, [O~\textsc{iii}] $\lambda5007$ can appear close to the wavelength expected for high-redshift \heii \citep[e.g.,][]{nagao_2008}. A foreground [O~\textsc{ii}]+[O~\textsc{iii}] emitter can therefore reproduce a two-line pattern. This interpretation is not possible for four of the eight candidates, for which $z_{\rm Ly\alpha}<3727/1215.67-1\simeq2.066$ and the observed \lya line lies blueward of rest-frame [O~\textsc{ii}] $\lambda3727$.

We test this foreground interpretation regardless by requiring a self-consistent low-redshift rest-frame optical line pattern. If the line identified as \lya were instead [O~\textsc{ii}], then [O~\textsc{iii}] $\lambda4959$, [O~\textsc{iii}] $\lambda5007$, and H$\beta$ should appear at predictable wavelengths. The [O~\textsc{iii}] doublet provides a direct quantitative test because [O~\textsc{iii}] $\lambda5007/\lambda4959=2.98$ \citep{storey_zeippen_2000}. If the stacked \heii feature were actually foreground [O~\textsc{iii}] $\lambda5007$, an accompanying [O~\textsc{iii}] $\lambda4959$ line should appear in the \heii-aligned stack at 1624.7~\AA\ with flux $\times 0.336$ of the apparent \heii line. We find no significant feature at this wavelength, measuring $L_{4959}=(0.046\pm0.147)\times10^{42}~{\rm erg~s^{-1}}$, and place a conservative $3\sigma$ limit of [O~\textsc{iii}] $\lambda4959$/\heii $<0.164$, below the $0.336$ ratio required if the stacked \heii feature were dominated by foreground [O~\textsc{iii}] $\lambda5007$.

Imaging and chance-alignment checks further disfavor the simplest foreground and coincident-LAE explanations; the full ancillary-data assessment is given in Appendix~\ref{app:ancillary_interlopers}.

Although individual candidates may still include low-redshift contaminants, this scenario does not naturally explain the stacked spectrum. The stack shows a coherent feature at the \heii position in the LAE rest frame without the accompanying low-redshift line pattern expected from a population of [\ion{O}{2}]+[\ion{O}{3}] interlopers. Rest-optical follow-up spectroscopy at the candidate redshifts will provide a direct test by measuring [\ion{O}{2}], H$\beta$, [\ion{O}{3}], and H$\alpha$.

\subsection{AGN Contamination}
\label{subsec:agn_contamination}

AGN are an obvious alternative because accretion-powered spectra can produce photons above the He$^{+}$ ionization threshold. In typical narrow-line regions, this hard spectrum is accompanied by UV permitted lines of high-ionization species, such as \civ and often \nv. Rest-frame UV line ratios thus provide a direct comparison to AGN-like ionization \citep[e.g.,][]{baldwin_1978,feltre_2016}.

The visual vetting already removes the most obvious AGN-like cases. In particular, candidates with strong broad \heii accompanied by clear \civ, \nv, or other UV metal lines are excluded from the final sample, even if their \heii feature is stronger than those of some of our final candidates. The AGN question for the final sample is therefore narrower: could a weak, obscured, low-luminosity, or otherwise unusual AGN produce strong \heii while keeping the UV metal lines below the observed stack limits?

Our stack does not show the UV line pattern expected for an ordinary metal-enriched AGN-dominated spectrum. We find $3\sigma$ limits of \civ/\mheii\ $<0.128$ and \nv/\mheii\ $<0.167$. These limits are far below typical narrow-line AGN UV ratios. For example, \citet{nagao_2005} report characteristic \civ/\mheii\ ratios of $1.42$ for high-redshift radio galaxies, 2.04 for type-II quasars, and 2.20 for Seyfert~2 galaxies. Our stacked limit is more than an order of magnitude lower than these values. The \heii width also disfavors an unobscured broad-line AGN interpretation: the stacked \heii FWHM is $559\pm84_{\rm stat}\pm7_{\rm sys}~\kms$, corresponding to an intrinsic width of $449~\kms$ (assuming an $R=900$ spectral resolution), far narrower than the broad permitted lines expected from broad-line AGN.

The available X-ray and radio data show no luminous counterpart, although they do not exclude a faint, obscured, or low-luminosity AGN; coverage details are provided in Appendix~\ref{app:ancillary_interlopers}.

A weak (or obscured) very metal-poor narrow-line AGN remains a possible explanation for some individual objects. Such a model would need to explain the presence of strong \heii while keeping \civ and \nv below the observed stack limits. The stacked line ratios and moderate intrinsic \heii width make a standard AGN-dominated explanation unlikely for the sample as a whole. Deeper rest-UV and rest-optical spectroscopy should test this scenario for individual candidates.

\subsection{Fast Shocks}
\label{subsec:shocks}

Fast radiative shocks can generate hard ionizing radiation and \heii emission \citep[e.g.,][]{dopita_sutherland_1996,allen_2008}. Shock and precursor models can also produce high-ionization UV metal lines, including \civ, \nv, and \oiii, with strengths that depend on the shock velocity, gas density, magnetic field, and metallicity.

The main challenge for a shock interpretation is the weakness of the metal lines. The stack shows \heii with EW$_{\rm rest}=28.7\pm5.6_{\rm stat}\pm3.7_{\rm sys}$~\AA, while \civ, \nv, and \oiii are constrained only as low upper limits relative to \heii. A shock model could suppress metal-line emission if the gas metallicity is extremely low, i.e., the shock interpretation requires the same low-metallicity conditions that motivate the stellar interpretation considered in Section~\ref{sec:physical_interpretation}. Shocks must also reproduce the observed \heii/\lya ratio and moderate intrinsic \heii width without creating stronger UV metal lines or clearly complex kinematic structure.

Shocks are not the most natural explanation for the stacked measurements. They remain possible in principle, especially in very low-metallicity gas, and deeper spectra could test for faint metal lines and broader wings. The current data do not show the metal-line signatures commonly expected from fast radiative shocks.

\subsection{Cooling Radiation}
\label{subsec:cooling_radiation}

Gravitational cooling radiation from gas accreting onto dark-matter halos can produce \lya and, in sufficiently hot low-metallicity gas, \heii \citep{yang_2006,scarlata_2009}. This channel is relevant since it can yield strong \lya and narrow \heii without significant metal-line emission. 

The main tension is quantitative. Simple cooling models predict \heii at least an order of magnitude fainter than \lya \citep{yang_2006}, whereas the stack has \mheii/Ly$\alpha=0.354\pm0.094$. This ratio alone is not decisive; suppression or redistribution of emergent Ly$\alpha$ would also increase \mheii/Ly$\alpha$. Morphologically, cooling-radiation interpretations are commonly associated with spatially extended \lya nebulae around gas condensing in massive dark-matter potential wells \citep[e.g.,][]{haiman_2000}. Seven of the eight sources do not satisfy the formal resolved-source criterion, and the mixed-sign individual velocity offsets further disfavor a single cooling or inflow interpretation. The spatial and kinematic tests are quantified in Appendices~\ref{app:lya_spatial} and \ref{app:lya_kinematics}. Cooling radiation remains possible for an individual object but is not favored as the dominant explanation for the stack.

\subsection{Wolf--Rayet Stars and Metal-enriched Stellar Populations}
\label{subsec:wr_stars}

Wolf--Rayet (WR) stars and related evolved massive-star populations can produce \heii emission, either as broad features from stellar winds or as narrow lines from the ionization caused by their UV hard stellar spectra \citep[e.g.,][]{schaerer_vacca_1998,brinchmann_2008,shirazi_brinchmann_2012}. These objects offer an important alternative to the very metal-poor or Pop~III-like interpretation.

A classical metal-enriched Wolf--Rayet interpretation is disfavored by the stacked line properties. The \heii emission produced by Wolf--Rayet populations is often broad ($\sim 1000$~\kms) \citep[e.g.,][]{wofford_2023}, whereas the stacked \heii line has an intrinsic FWHM of $449~\kms$. Moreover, Wolf--Rayet stars in metal-rich populations are also expected to show features associated with evolved massive stars and their surrounding gas. The HETDEX stacked spectrum instead shows weak or undetected \civ, \nv, and \oiii.

These constraints do not rule out stellar photoionization. The CLASSY results reinforce this point at low redshift, showing that \heii can arise in local star-forming analogs yet remains embedded within a broader UV line pattern that ultimately carries the diagnostic power \citep{berg_2022,mingozzi_2022}. The physical origin of narrow nebular \heii in star-forming galaxies remains debated \citep[e.g.,][]{brinchmann_2008,grafener_2015}. Importantly, \citet{grafener_2015} reports that very massive stars at sub-solar metallicity can develop slow, dense winds that produce narrow He~\textsc{ii} $\lambda1640$ emission with an intrinsic FWHM of $\sim300$--500~\kms, overlapping the intrinsic width measured in our stack ($449\pm105~\kms$). A low FWHM alone is therefore insufficient to exclude a metal-poor very massive star (VMS) interpretation. The stronger constraint arises from the UV metal-line limits: even populations with a metal abundance of a few percent solar are expected to produce observable \civ, \nv, or \oiii emission. Our stack therefore suggests much lower abundances. The data primarily disfavor standard metal-enriched Wolf--Rayet and VMS scenarios; while a very metal-poor VMS population cannot be excluded, it is physically close to the Pop~III-like regime discussed in Section~\ref{sec:physical_interpretation}.

\subsection{Summary of Contaminant Tests}
\label{subsec:contaminant_summary}

Table~\ref{tab:contaminant_assessment} summarizes the contaminant assessment. The alternatives differ in detail, but the same tension recurs: while several processes can produce emission coincident with \heii, their physics also produce artifacts, spatial inconsistencies, broad lines, foreground line patterns, or UV metal emission that is stronger than observed in the final vetted sample and its stack. Our tests do not conclusively eliminate every alternative for every object, but they demonstrate that the final sample is difficult to reproduce with standard contaminant scenarios.

\begin{table*}
\centering
\caption{
Summary of contaminant assessment. The assessments apply to the eight-object stack; they do not rule out every scenario for every individual candidate.
}
\label{tab:contaminant_assessment}
\footnotesize
\setlength{\tabcolsep}{3pt}
\renewcommand{\arraystretch}{1.25}

\begin{tabular}{llll}
\hline
\hline
Scenario & Expected Signature & Observed Behavior & Assessment \\
\hline

\tcell{0.13\textwidth}{Noise/artifacts} &
\tcell{0.25\textwidth}{Isolated spikes, skyline residuals, wavelength-edge effects, spatial mismatches, or comparable signals in arbitrary nearby windows.} &
\tcell{0.29\textwidth}{Final candidates pass visual, fiber-level, imaging, metal-line, AGN, and interloper checks; signal appears in both \lya- and \heii-aligned stacks; blank windows in the same spectra do not reproduce a comparable positive feature.} &
\tcell{0.20\textwidth}{Highly unlikely for the vetted stack.} \\[0.8em]

\tcell{0.13\textwidth}{Low-$z$ interloper} &
\tcell{0.25\textwidth}{[O~\textsc{ii}]+[O~\textsc{iii}]+H$\beta$ pattern; possible low-$z$ imaging counterpart.} &
\tcell{0.29\textwidth}{Stack does not show the expected low-$z$ line pattern, including the required [O~\textsc{iii}] $\lambda4959$ counterpart; no obvious bright foreground counterpart in the extraction.} &
\tcell{0.20\textwidth}{Possible for individual objects; very unlikely for the stack.} \\[0.8em]

\tcell{0.13\textwidth}{Broad-line AGN} &
\tcell{0.25\textwidth}{Very broad permitted lines; strong \civ and/or \nv.} &
\tcell{0.29\textwidth}{Approximate intrinsic \heii FWHM $=449~{\rm km~s^{-1}}$; \civ/\mheii\ $<0.128$ and \nv/\mheii\ $<0.167$.} &
\tcell{0.20\textwidth}{Strongly disfavored.} \\[0.8em]

\tcell{0.13\textwidth}{Narrow-line AGN} &
\tcell{0.25\textwidth}{Narrow \heii with strong high-ionization UV metal lines.} &
\tcell{0.29\textwidth}{Metal-line-rich candidates rejected in vetting; stack limits are \civ/\mheii\ $<0.128$ and \nv/\mheii\ $<0.167$, more than $10\times$ below typical AGN values.} &

\tcell{0.20\textwidth}{Possible for individual objects; unlikely for the stack.} \\[0.8em]

\tcell{0.13\textwidth}{Fast shocks} &
\tcell{0.25\textwidth}{\heii with \civ, \nv, \oiii, and complex kinematics.} &
\tcell{0.29\textwidth}{Strong \heii with weak metal-line limits and moderate intrinsic line width.} &
\tcell{0.20\textwidth}{Disfavored unless the gas is very metal poor.} \\[0.8em]

\tcell{0.13\textwidth}{Cooling radiation} &
\tcell{0.25\textwidth}{Extended \lya nebulae; \heii expected $\gtrsim10\times$ fainter than \lya; weak \civ and \nv.} &
\tcell{0.29\textwidth}{The stack has \heii/\lya$=0.354\pm0.094$ (\lya/\mheii\ $\simeq2.8$). Only one of the eight candidates is formally resolved in Ly$\alpha$.} &
\tcell{0.20\textwidth}{Possible for individual objects; disfavored for the stack.} \\[0.8em]

\tcell{0.13\textwidth}{Wolf--Rayet stars} &
\tcell{0.25\textwidth}{Broad stellar \heii and metal-enriched wind or nebular features.} &
\tcell{0.29\textwidth}{Moderate intrinsic \heii width; weak \civ, \nv, and \oiii.} &
\tcell{0.20\textwidth}{Classical metal-enriched WR interpretation disfavored.} \\

\noalign{\vskip 2pt}
\hline
\end{tabular}
\end{table*}

\section{Physical Interpretation}
\label{sec:physical_interpretation}

The HETDEX stack reveals the rare rest-frame UV line pattern targeted by the selection: strong \heii, high \heii/\lya, moderate intrinsic \heii line width, and $3\sigma$ upper limits on \nv, \civ, and \oiii\ that are well below the levels expected from standard contaminants. Section~\ref{sec:contaminant_tests} shows that none of these scenarios naturally reproduce this combination. This section identifies the most plausible physical origin.

\subsection{Why the Line Pattern Points to Hard Ionizing Radiation in Low-metallicity Gas} 
\label{subsec:why_extreme} 

The unusual aspect of the stacked spectrum is not \heii\ emission alone but the absence of UV metal-line emission. Photons above 54.4~eV are required for \heii\ recombination, but AGN and fast shocks that supply such photons normally also produce \civ, \nv, and \oiii\null. Similarly, classical Wolf--Rayet populations produce \heii\ with broad stellar-wind features in metal-enriched gas. The stacked measurements therefore require either a hard ionizing source in very metal-poor gas, where collisionally excited metal-line emission is naturally suppressed, or an unusual mechanism that produces \heii\ while circumventing the standard metal-line diagnostics. The rarity of the sample is in line with the scarcity of extreme ionization conditions at $z\sim2$.

\subsection{Stellar Population Interpretation}
\label{subsec:stellar_population_expectations}

Normal metal-enriched stellar populations generally produce too few photons above 54.4~eV to explain strong nebular \heii at the EW measured in the stack. Binary evolution, stripped stars, very massive stars, X-ray binaries, and Wolf--Rayet stars can harden the ionizing spectrum and may produce nebular \heii under some conditions \citep[e.g.,][]{schaerer_2003,raiter_2010,shirazi_brinchmann_2012,senchyna_2020}. These channels remain relevant alternatives, especially for individual candidates. However, such an interpretation in a metal-rich population must also explain why \heii is strong while \nv, \civ, and \oiii remain weak in the stack.

Very metal-poor stellar populations provide a more direct route to the observed line pattern. Low-metallicity massive stars are hotter and produce harder ionizing spectra, while extremely low metal abundances limit the number of metal ions available to produce collisionally excited UV lines. This relation is not strictly monotonic: weaker metal cooling can raise the electron temperature, but at very low abundance the small number of metal ions still makes metal-line emission weak. Strong \heii with weak \civ, \nv, and \oiii is therefore qualitatively consistent with stellar photoionization in extremely metal-poor gas.

A Pop~III-like ionizing spectrum is the most extreme version of this interpretation. Models of very massive, metal-free or nearly metal-free stars predict hard ionizing spectra, strong \heii, strong hydrogen recombination lines, and large rest-frame EWs \citep[e.g.,][]{tumlinson_shull_2000,schaerer_2003,raiter_2010,nakajima_maiolino_2022,venditti_2026}. Such phases should be short-lived because the most massive stars evolve quite rapidly, so active Pop~III-like systems are expected to be rare.

At $z\sim2$, a purely pristine stellar population would be surprising, but it remains physically plausible for extremely low-metallicity star formation to occur if metal enrichment and mixing are spatially inhomogeneous. Simulations and semi-analytic models suggest that residual Pop~III or extremely metal-poor star formation can persist after $z\sim10$ in chemically young gas pockets, especially in low-mass or inefficiently enriched environments \citep[e.g.,][]{tornatore_2007,johnson_2013,sarmento_2018,venditti_2023}.

The stacked \heii luminosity, $L_{\rm HeII}=(2.98\pm0.38)\times10^{42}~{\rm erg\,s^{-1}}$, is at the bright end of the range predicted by Pop~III photoionization models. \citet{venditti_2026} finds that \heii luminosities up to ${\sim}5$--$6\times10^{42}$~erg~s$^{-1}$ are achievable for Pop~III systems with very top-heavy initial mass functions (IMFs) and total cluster masses ${\sim}6\times10^5\,M_\odot$. Such masses are not expected in typical dark-matter minihalos; they likely require more extreme environments such as strongly irradiated atomic-cooling halos in which H$_2$ dissociation by a Lyman-Werner background delays fragmentation \citep[e.g.,][]{haiman_1997} and allows the accumulation of larger protostellar masses \citep[e.g.,][]{greif_bromm_2006,greif_2008,bromm_2009,sugimura_2023,jeong_2026}. Additional luminosity boosting is possible if the Pop~III stars undergo chemically homogeneous evolution (CHE), which is expected for rapid rotators. This hardens the ionizing spectrum and may prolong the \mheii-bright phase \citep[e.g.,][]{sibony_2022,wasserman_2026}. Taken at face value, the stacked luminosity is not impossible for a Pop~III-like system, but the scenario presents challenges for a typical minihalo cluster, and implies either a large total Pop~III mass, a top-heavy IMF, a favorable host environment, or a contribution from multiple unresolved systems within the HETDEX aperture. We caution that the stacked \heii luminosity is the average over eight objects whose individual luminosities are not well constrained; some may be significantly brighter or fainter. A future clustering or host-halo-mass analysis could test whether the environments of the HETDEX candidates are consistent with the more extreme conditions required for bright Pop~III-like \heii emission.

\subsection{Comparison to Photoionization Model Diagnostics}
\label{subsec:cloudy_grids}

Figure~\ref{fig:heii_diagnostic_grid} compares the stacked measurements with the photoionization models of \citet{nakajima_maiolino_2022}. This comparison is interpretive rather than definitive because the predicted ratios depend on the ionizing spectrum, gas metallicity, ionization parameter, density, geometry, stellar-population age, and  IMF.  The closest model with a He~\textsc{ii} EW consistent with the observed value is a dust-included $Z=10^{-4}$ model with a $1$--$500\,M_\odot$ IMF. While the model closely reproduces the observed He~\textsc{ii} EW, the observed He~\textsc{ii}/Ly$\alpha$ ratio is approximately $5.4$ times higher.

\citet{nakajima_maiolino_2022} notes that the \texttt{Cloudy} Ly$\alpha$ luminosities represent approximately the maximum emission emerging from the ionized cloud. Attenuation or redistribution by neutral gas, dust, and Ly$\alpha$ radiative transfer would weaken the emergent Ly$\alpha$ emission and increase He~\textsc{ii}/Ly$\alpha$ above the model values. Dust can also modestly increase the He~\textsc{ii} EW by reducing the UV continuum. The model He~\textsc{ii}/Ly$\alpha$ ratios should therefore be regarded as lower limits. If the difference from the closest EW-compatible model were attributed entirely to attenuation or redistribution of Ly$\alpha$ at fixed He~\textsc{ii} emission, it would correspond to an effective Ly$\alpha$ reduction of approximately $80\%$ relative to the model prediction. The independently large He~\textsc{ii} EW demonstrates that genuinely strong He~\textsc{ii} emission contributes to the observed ratio, while Ly$\alpha$ suppression or redistribution likely also makes a substantial contribution.

Taken together, the diagnostic panels provide stronger constraints than any individual ratio. The stack combines a large \heii EW and high \heii/\lya with stringent upper limits on \civ/\mheii, \oiii/\mheii, and \nv/\mheii. These limits disfavor the strong UV metal-line emission commonly associated with AGN-like ionization and metal-enriched stellar populations, while remaining consistent with a very hard ionizing spectrum in very low-metallicity gas. The comparison does not uniquely identify the stellar population: the measurements remain compatible with an extremely metal-poor Pop~II population, a short-lived Pop~III dominated burst, or another hard ionizing source embedded in very low-metallicity gas.

\subsection{Ly$\alpha$ Kinematics and Radiative-transfer Implications}
\label{subsec:lya_kinematic_interpretation}

The individual \lya--\heii offsets are heterogeneous. Three candidates are significantly redshifted, three are significantly blueshifted, and two are consistent with zero. The \heii-aligned stack offset is also consistent with zero. A single outflow, inflow, or neutral-gas configuration therefore cannot account for all eight line profiles.

Positive Ly$\alpha$ offsets can arise from resonant transfer through outflowing or high-column neutral gas, but the offset does not determine a unique H~\textsc{i} column because covering fraction and geometry are degenerate. Negative offsets can be produced by complex or multi-component radiative transfer, inflowing material, or cooling-related emission; low-S/N centroid shifts provide an additional observational degeneracy. The present measurements therefore establish object-to-object kinematic diversity and no unique flow direction. Higher S/N \heii or rest-optical optically thin lines are required to establish more precise systemic redshifts for the individual candidates.

\subsection{Comparison to Previous Observational \& Theoretical Constraints}
\label{subsec:previous_constraints}

The number density of ${\sim}30~\mathrm{Gpc}^{-3}$ provides a rough comparison point for previous Pop~III and very metal-poor searches. Earlier \mheii-selected surveys at $z\sim2$--4 place limits on similar short-lived, metal-line-poor phases \citep{nagao_2008,cassata_2013}, while recent JWST studies report candidates at higher redshift \citep[e.g.,][]{fujimoto_2025,morishita_2025}. As a plausibility check, if each candidate hosts $M_{\rm Pop,III}\sim6\times10^5\,M_\odot$ of Pop~III stars \citep{venditti_2026}, then $n_{\rm cand}\sim30~{\rm Gpc}^{-3}$ implies a Pop~III stellar-mass density of ${\sim}0.018\,M_\odot\,{\rm Mpc}^{-3}$. For a characteristic \mheii\ burst visibility time of $t \sim3$~Myr \citep{schaerer_2002}, this measurement corresponds to a crude candidate star-formation rate density (SFRD) of $6\times10^{-9}\,M_\odot\,{\rm yr}^{-1}\,{\rm Mpc}^{-3}$.
This value lies below the \citet{nagao_2008} upper limit of $\lesssim5\times10^{-6}\,M_\odot\,{\rm yr}^{-1}\,{\rm Mpc}^{-3}$, below the \citet{cassata_2013} estimate of $\sim10^{-6}\,M_\odot\,{\rm yr}^{-1}\,{\rm Mpc}^{-3}$, and far below JWST-based estimates at $z\sim6$--7 \citep[e.g.,][]{fujimoto_2025}. It also falls below theoretical expectations for late Pop~III star formation in inefficiently mixed gas, which predict SFRDs of order $10^{-5}$--$10^{-6}\,M_\odot\,{\rm yr}^{-1}\,{\rm Mpc}^{-3}$ at comparable redshifts \citep{tornatore_2007,liu_bromm_2020}. The low rate is expected for a deliberately restrictive search at $z\sim2$, and should be interpreted as an observed candidate SFRD for the rare \mheii-bright, metal-line-poor phase, not a completeness-corrected census of all Pop~III systems.

\subsection{Final Verdict}
\label{subsec:main_interpretation}

The most natural physical interpretation of the stacked line pattern is a hard ionizing source in low-metallicity gas. A very metal-poor stellar component, possibly including a Pop~III-like contribution, is consistent with the measured \heii\ EW, \heii/\lya line ratio, the intrinsic \mheii\ line width, and UV metal-line limits. The present data do not directly measure gas-phase metallicity, cannot confirm a Pop~III origin, and cannot exclude all alternatives for individual candidates. Rest-optical spectroscopy of \Hb, \OIII~$\lambda4959,\lambda5007$, \OII~$\lambda3727$, and \Ha\ is required to constrain metallicity, ionization parameter, and AGN diagnostics before any stronger conclusion can be drawn.

\section{Future Investigations} 
\label{sec:future_followup} 

Three spectroscopic steps will determine whether the eight candidates are genuinely very metal-poor systems. 

\textit{Deeper rest-frame UV spectroscopy.} Higher-S/N optical spectra covering the observed-frame \lya, \mheii, \nv, \civ, and \oiii\ regions are the immediate priority. Deeper UV spectra will test whether the weak-metal-line pattern persists in individual targets, improve \lya\ and \heii\ line-profile measurements, place tighter limits on UV metal lines, and identify any faint broad-line AGN features or spectral artifacts unresolved in the present HETDEX data. Deeper UV spectra will also reveal object-to-object variation that the current stack cannot resolve.

\textit{Rest-frame optical spectroscopy.} At $z\sim2$, \OII\ $\lambda3727$, \Hb, \OIII~$\lambda\lambda4959,5007$, and \Ha\ all fall in the near-infrared. These lines measure gas-phase metallicity, provide Balmer-line diagnostics of the ionizing spectrum, dust content, and star-formation rate, and enable tests for AGN activity. The observations would also directly test the low-redshift interloper interpretation: a genuine LAE at $z\sim2$ should not show the [O~\textsc{ii}]+[O~\textsc{iii}]+H$\beta$ pattern expected from a foreground emitter. Ground-based near-infrared spectrographs can reach the brightest candidates efficiently and should precede JWST observations. 

\textit{JWST/NIRSpec observations.} If candidates are confirmed to have weak metal lines and no AGN signature, the next step is JWST Near Infrared Spectrograph (NIRSpec) spectroscopy. NIRSpec can tighten the metallicity constraints and, through high-S/N spectroscopy, distinguish very metal-poor stellar photoionization from conventional contaminants. Supporting photometric constraints on stellar mass and star-formation rate are useful but secondary to spectroscopic confirmation. 

The highest-priority targets for each step are those that pass the previous one: confirmed presence of rest-UV \heii, stringent limits on metal-line strengths, and the absence of AGN signature lines define the candidates most worth investing in at each successive stage. Additionally, future analysis of the large-scale clustering or host-halo properties of the HETDEX \heii candidates would test whether the objects occupy the more extreme environments that would be required if the bright stacked \heii luminosity reflects a genuine Pop~III-like source.

\section{Conclusions}
\label{sec:conclusions}

We searched 109,545 high-confidence LAEs at $1.9 < z < 2.3$ in HETDEX for systems with strong He~\textsc{ii} $\lambda1640$ emission and no detected UV metal-line emission. Eight candidates survive automated selection and visual vetting. The main results are as follows.

\begin{itemize}
    \item \textbf{The candidate objects are rare and the selection is purity-oriented (\S\S~\ref{subsec:final_candidate_sample} \& \ref{subsec:selection_rarity}).} Eight candidates from 109,545 LAEs correspond to an observed yield of $7.3\times10^{-5}$, equivalent to a candidate comoving number density of ${\sim}30~\mathrm{Gpc}^{-3}$. This value is not a completeness-corrected occurrence rate for all Pop~III-active or very metal-poor systems; the selection intentionally excludes hybrid, self-enriched, and metal-line-rich \heii emitters.
    
    \item \textbf{The stacked He~\textsc{II} signal is highly significant (\S\S~\ref{subsec:stacked_spectrum_all8} \& \ref{subsec:measurement_uncertainties}).} The eight-object stack has \mheii\ EW$_{\rm rest}=28.7\pm6.7$~\AA, \mheii/Ly$\alpha = 0.354\pm0.094$, and an intrinsic He~\textsc{ii} FWHM $= 449\pm105~\kms$. The measurement is stable to leave-one-out removal and choice of stacking method, and is not dominated by any single object.

    \item \textbf{UV metal lines are undetected in the stack (\S\S~\ref{subsec:metal_upper_limits} \& \ref{subsec:ew_line_ratios}).} We place $3\sigma$ upper limits of \nv/\mheii\ $< 0.167$, \civ/\mheii\ $< 0.128$, and \oiii/\mheii\ $< 0.140$. The stacked spectrum shows no significant flux in any of the three primary UV metal-line diagnostics used to identify AGN, shocks, and metal-enriched stellar populations.

    \item \textbf{The Ly$\alpha$ kinematics and spatial properties are heterogeneous (\S\S~\ref{subsec:individual_candidates}, \ref{subsec:stacked_spectrum_all8}).} The individual Ly$\alpha$--He~\textsc{ii} velocity offsets include three significant positive values, three significant negative values, and two values consistent with zero, while the He~\textsc{ii}-aligned stack gives $\Delta v_{\rm Ly\alpha-HeII}=+13^{+264}_{-173}~\kms$, consistent with zero. One candidate is formally resolved in Ly$\alpha$; the remaining seven do not satisfy the complete resolved-source criterion.

    \item \textbf{Standard contaminants are disfavored (\S~\ref{sec:contaminant_tests} \& Table \ref{tab:contaminant_assessment}).} The moderate intrinsic He~\textsc{ii} width disfavors broad-line AGN. The weak UV metal-line limits are inconsistent with ordinary narrow-line AGN, fast radiative shocks, and classical Wolf--Rayet populations, which all deposit significant flux in at least one of \nv, \civ, or \oiii. Individual objects may have contamination, but no single standard mechanism explains the stack.

    \item \textbf{A very metal-poor or Pop~III-like ionizing source is the most natural interpretation (\S~\ref{subsec:main_interpretation} \& Figure \ref{fig:heii_diagnostic_grid}).} Hard ionizing radiation in low-metallicity gas reproduces the observed He~\textsc{ii} strength and UV metal-line weakness. The current data cannot confirm a Pop~III origin, measure gas-phase metallicity, or rule out all alternatives; they identify eight objects worth testing at greater depth.
    
\end{itemize}

The immediate priority is deeper rest-frame UV spectroscopy to measure individual He~\textsc{ii} detections and test whether the weak-metal-line pattern persists object by object. For candidates with confirmed He~\textsc{ii} and no detected UV metal lines, rest-optical spectroscopy of H$\beta$, [O~\textsc{iii}] $\lambda4959,5007$, [O~\textsc{ii}] $\lambda3727$, and H$\alpha$ will then constrain gas-phase metallicity, ionization conditions, and AGN activity. This observational sequence will determine whether these candidates have a Pop~III origin.

\acknowledgments
We thank the referee for their insightful feedback and constructive suggestions, which improved the clarity and quality of the paper.

HETDEX is led by the University of Texas at Austin McDonald Observatory and Department of Astronomy with participation from the Ludwig-Maximilians-Universit\"at M\"unchen, Max-Planck-Institut f\"ur Extraterrestrische Physik (MPE), Leibniz-Institut f\"ur Astrophysik Potsdam (AIP), Texas A\&M University, The Pennsylvania State University, Institut f\"ur Astrophysik G\"ottingen, The University of Oxford, Max-Planck-Institut f\"ur Astrophysik (MPA), The University of Tokyo, and Missouri University of Science and Technology. In addition to Institutional support, HETDEX is funded by the National Science Foundation (grant AST-0926815), the State of Texas, the US Air Force (AFRL FA9451-04-2-0355), and generous support from private individuals and foundations.

Observations were obtained with the Hobby-Eberly Telescope (HET), which is a joint project of the University of Texas at Austin, the Pennsylvania State University, Ludwig-Maximilians-Universit\"at M\"unchen, and Georg-August-Universit\"at G\"ottingen. The HET is named in honor of its principal benefactors, William P. Hobby and Robert E. Eberly.

VIRUS is a joint project of the University of Texas at Austin, Leibniz-Institut f\"ur Astrophysik Potsdam (AIP), Texas A\&M University (TAMU), Max-Planck-Institut f\"ur Extraterrestrische Physik (MPE), Ludwig-Maximilians-Universit\"at M\"unchen, Pennsylvania State University, Institut f\"ur Astrophysik G\"ottingen, University of Oxford, and the Max-Planck-Institut f\"ur Astrophysik (MPA). In addition to Institutional support, VIRUS was partially funded by the National Science Foundation, the State of Texas, and generous support from private individuals and foundations.

The authors acknowledge the Texas Advanced Computing Center (TACC) at The University of Texas at Austin for providing high performance computing, visualization, and storage resources that have contributed to the research results reported within this paper. URL: http://www.tacc.utexas.edu

The Institute for Gravitation and the Cosmos is supported by the Eberly College of Science and the Office of the Senior Vice President for Research at the Pennsylvania State University.

KG acknowledges support from NSF-2008793.
JBM acknowledges support from NSF Grants AST-2307354 and AST-2408637, and NASA through grant JWST-GO-03224. 
SS acknowledges support from the National Science Foundation under grants NSF-2219212 and NSF-2511145.
RC and CG acknowledge support from the National Science Foundation under grant AST-2408358.

\facility{HET}

\software{Astropy \citep{astropy:2018}, NumPy \citep{numpy}, SciPy \citep{SciPy}, Matplotlib \citep{matplotlib}, ELiXer \citep{davis_hetdex_2023}, emcee \citep{foremanmackey_2013}}

\newpage
\bibliography{HeII}

\appendix

\section{Candidate-validation Details}
\label{app:candidate_validation}

\subsection{Fiber-level and Extraction-center Checks}
\label{app:fiber_checks}

For each candidate, we use the PSF-weighted spectrum extracted at the LAE position and inspect the contributing fiber spectra around both \lya and \heii. A plausible \lya+\heii source should show both lines at a common sky position; emission in spatially distinct fibers would instead suggest a foreground source, blend, or localized artifact. We give particular weight to features appearing in fibers within $1\farcs5$ of the extraction center, comparable to the VIRUS fiber diameter. 

The fiber-level information provides a check independent of the one-dimensional Gaussian fit. A detector artifact or random noise spike can appear in a single fiber, whereas a real source should remain consistent between the fibers nearest the extraction center. We find no final candidate in which \lya appears in one spatially distinct fiber while candidate \heii appears only in another unrelated fiber. We also test small shifts in extraction center and find no strongly displaced \heii signal. These tests establish positional consistency at the spatial precision of the data; they are not measurements of a resolved \heii size.

Figure~\ref{fig:example_candidate_3005020033} shows the procedure for one representative source. Its full spectrum places candidate \heii near the wavelength predicted from the \lya-based redshift, the velocity panels display the two line regions, and the imaging and fiber overlays test for a nearby continuum source or a spatially distinct line origin. This example illustrates the validation logic rather than assigning quantitative weight to an individual-object line ratio. A full grid search over extraction center is beyond the scope of this paper and we defer it to a future paper.

\subsection{Spectral and Fiber-level Checks for the Remaining Candidates}
\label{app:remaining_candidate_checks}

Figure~\ref{fig:remaining_candidate_checks} shows the spectral and
fiber-level diagnostics for the remaining seven final candidates. For each candidate, the
PSF-weighted rest-frame spectrum is shown together with the observed-frame
spectra of the four nearest fibers. These panels provide qualitative checks for
spatial consistency and localized artifacts.

\newcommand{\candidatepanel}[2]{%
  \IfFileExists{figures/#2}{%
    \includegraphics[width=0.46\textwidth]{figures/#2}%
  }{%
    \fbox{\parbox[c][1.35in][c]{0.43\textwidth}{\centering #1}}%
  }%
}

\begin{figure*}[t]
\centering
\subfloat[detectid 3007998098.]{%
  \candidatepanel{detectid 3007998098}{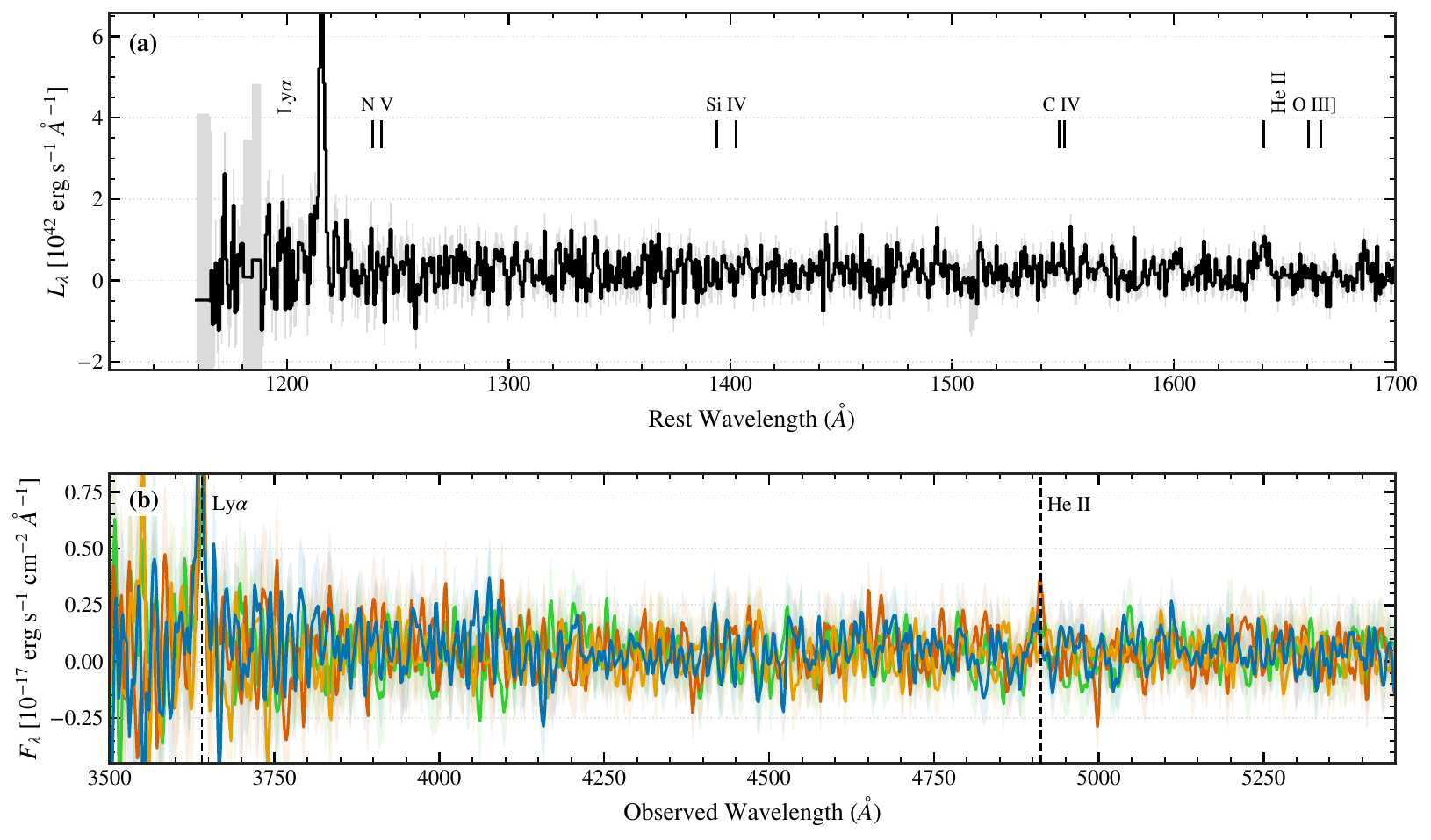}}
\hfill
\subfloat[detectid 3009751099.]{%
  \candidatepanel{detectid 3009751099}{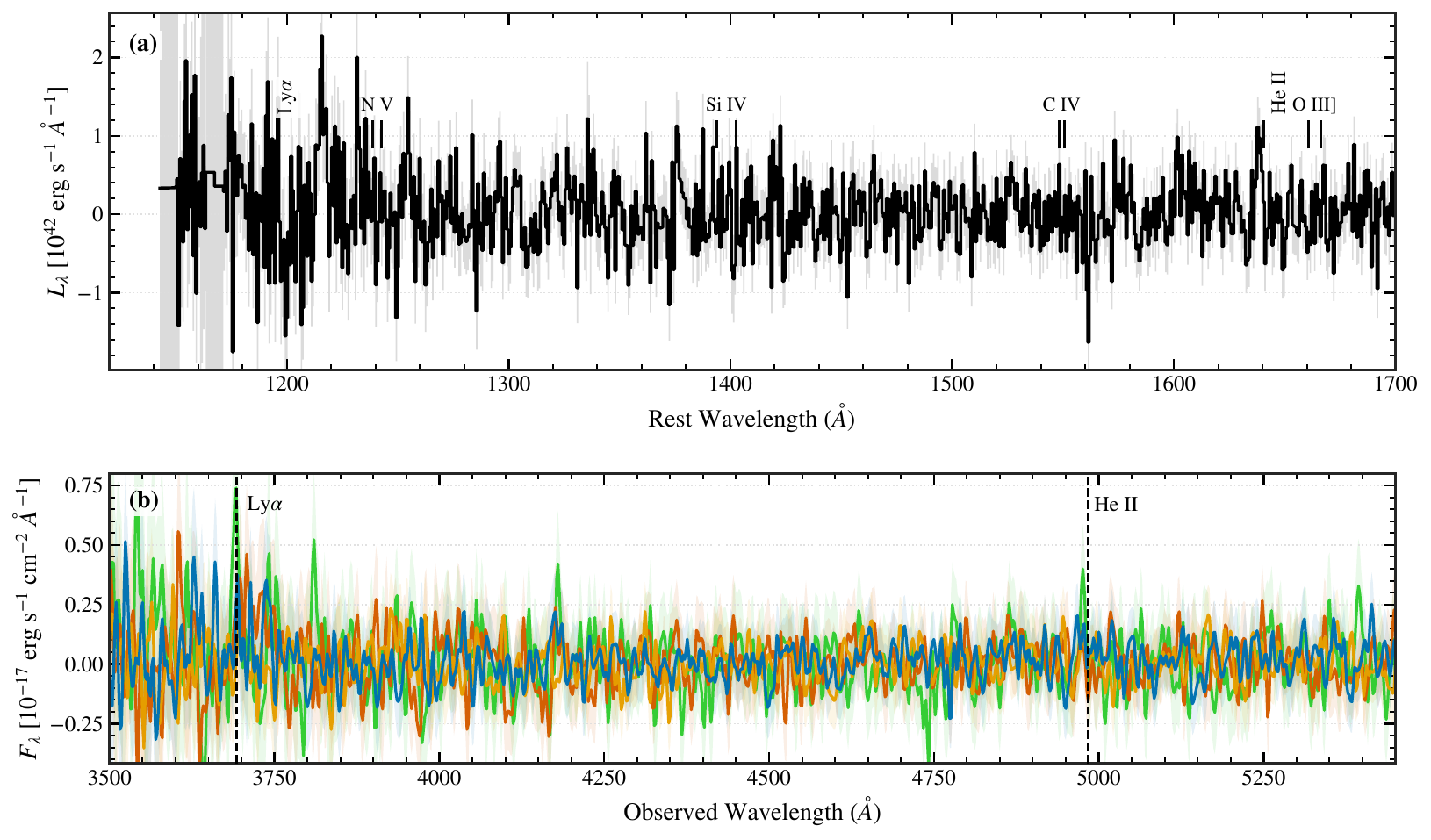}}\\[6pt]
\subfloat[detectid 3010547400.]{%
  \candidatepanel{detectid 3010547400}{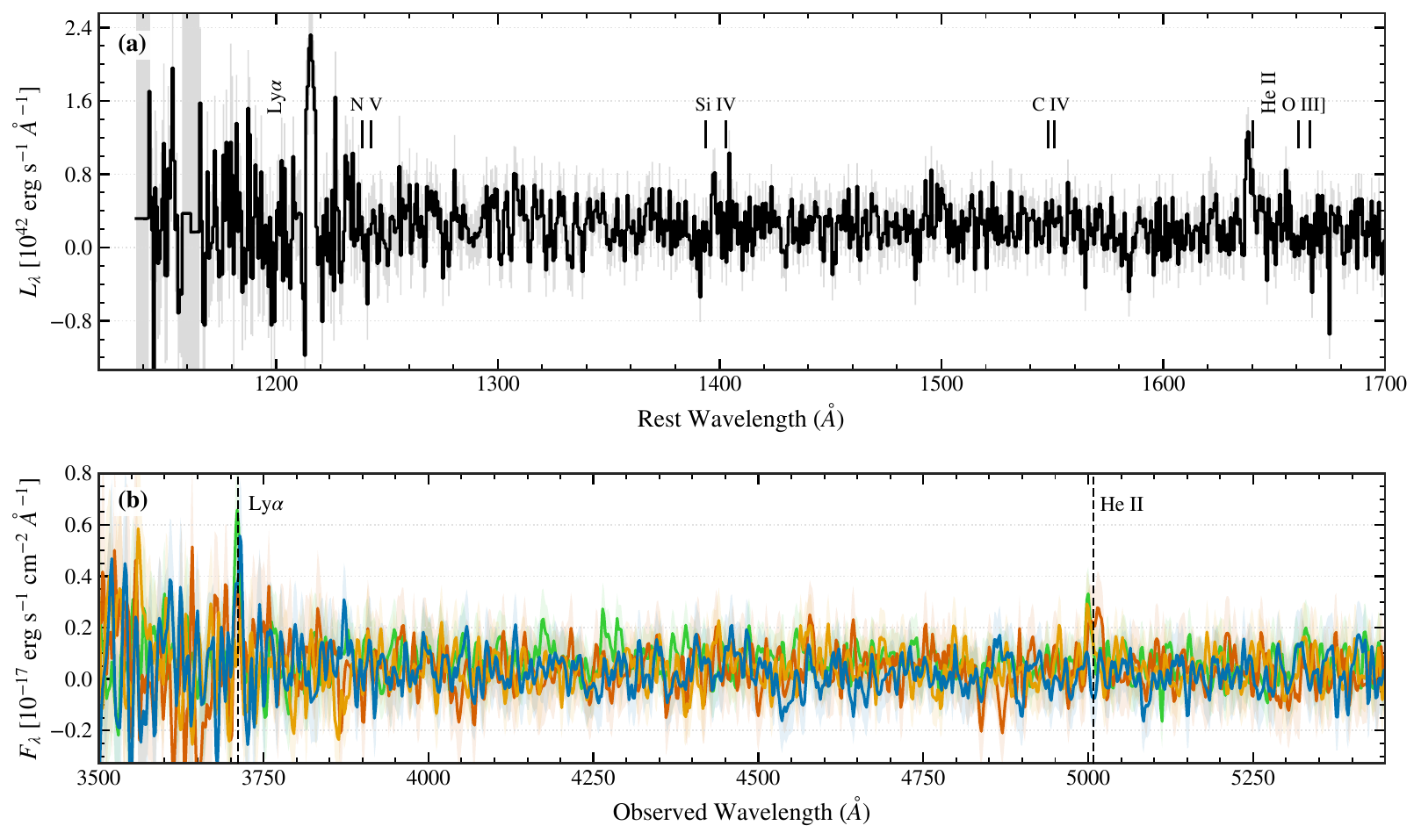}}
\hfill
\subfloat[detectid 4019081078.]{%
  \candidatepanel{detectid 4019081078}{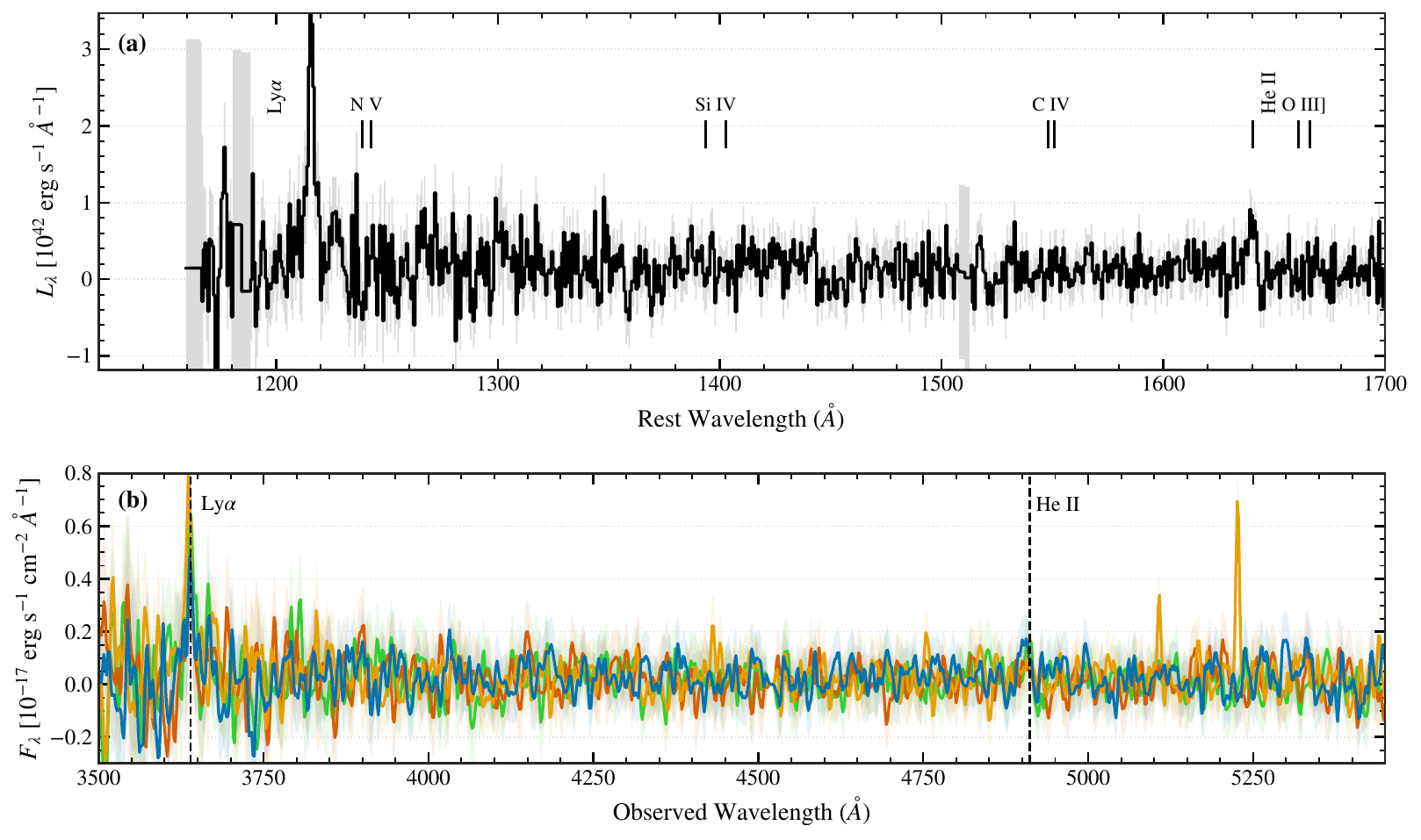}}\\[6pt]
\caption{\textbf{Spectral and fiber-level checks for the seven final
candidates not shown in Figure~\ref{fig:example_candidate_3005020033}.}
Each subfigure presents the PSF-weighted rest-frame UV spectrum and the
observed-frame spectra of the four nearest fibers. The fiber colors follow the
same nearest-to-farthest ordering used in
Figure~\ref{fig:example_candidate_3005020033}.}
\label{fig:remaining_candidate_checks}
\end{figure*}

\begin{figure*}[t]
\ContinuedFloat
\centering
\subfloat[detectid 4028177445.]{%
  \candidatepanel{detectid 4028177445}{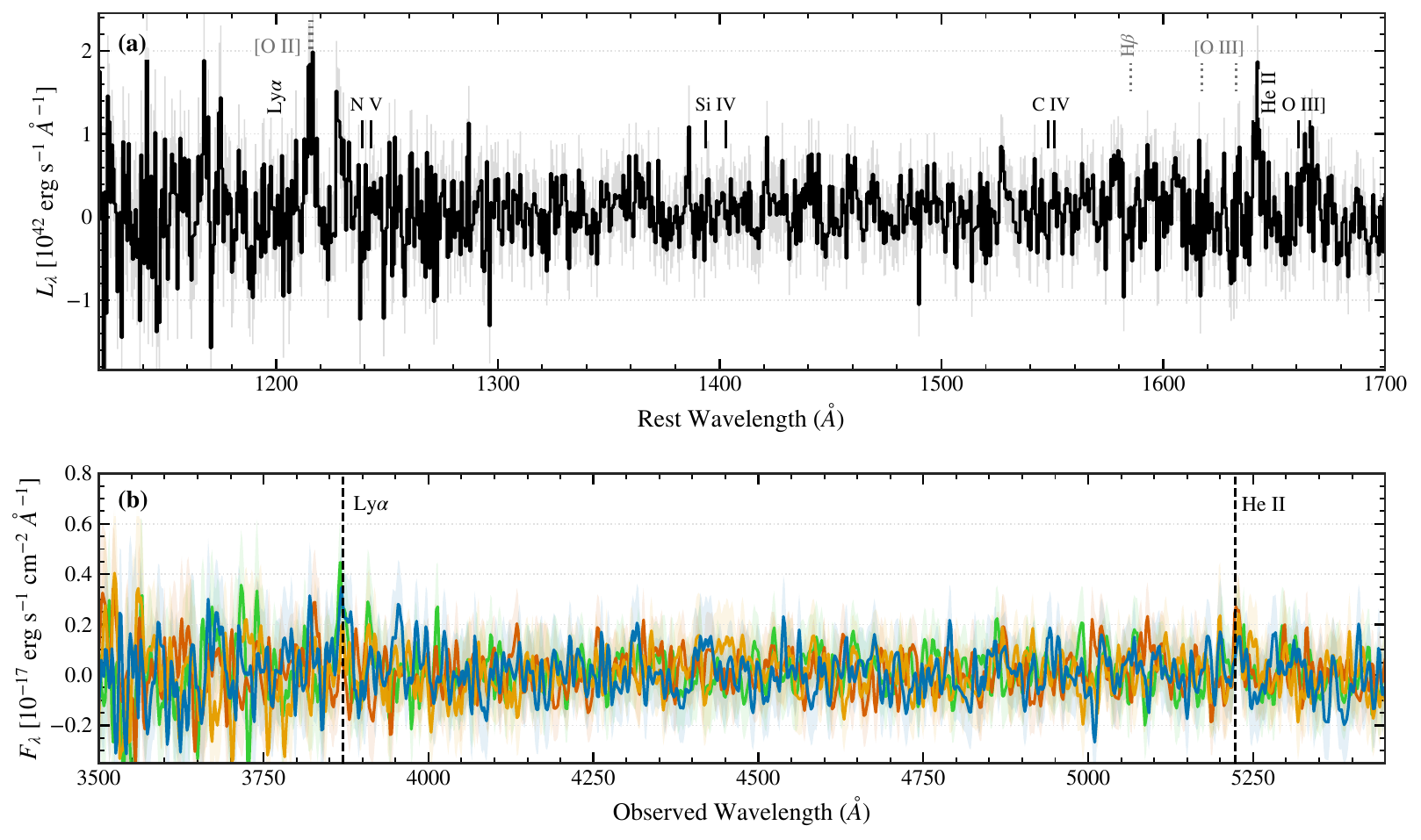}}
\hfill
\subfloat[detectid 4028632963.]{%
  \candidatepanel{detectid 4028632963}{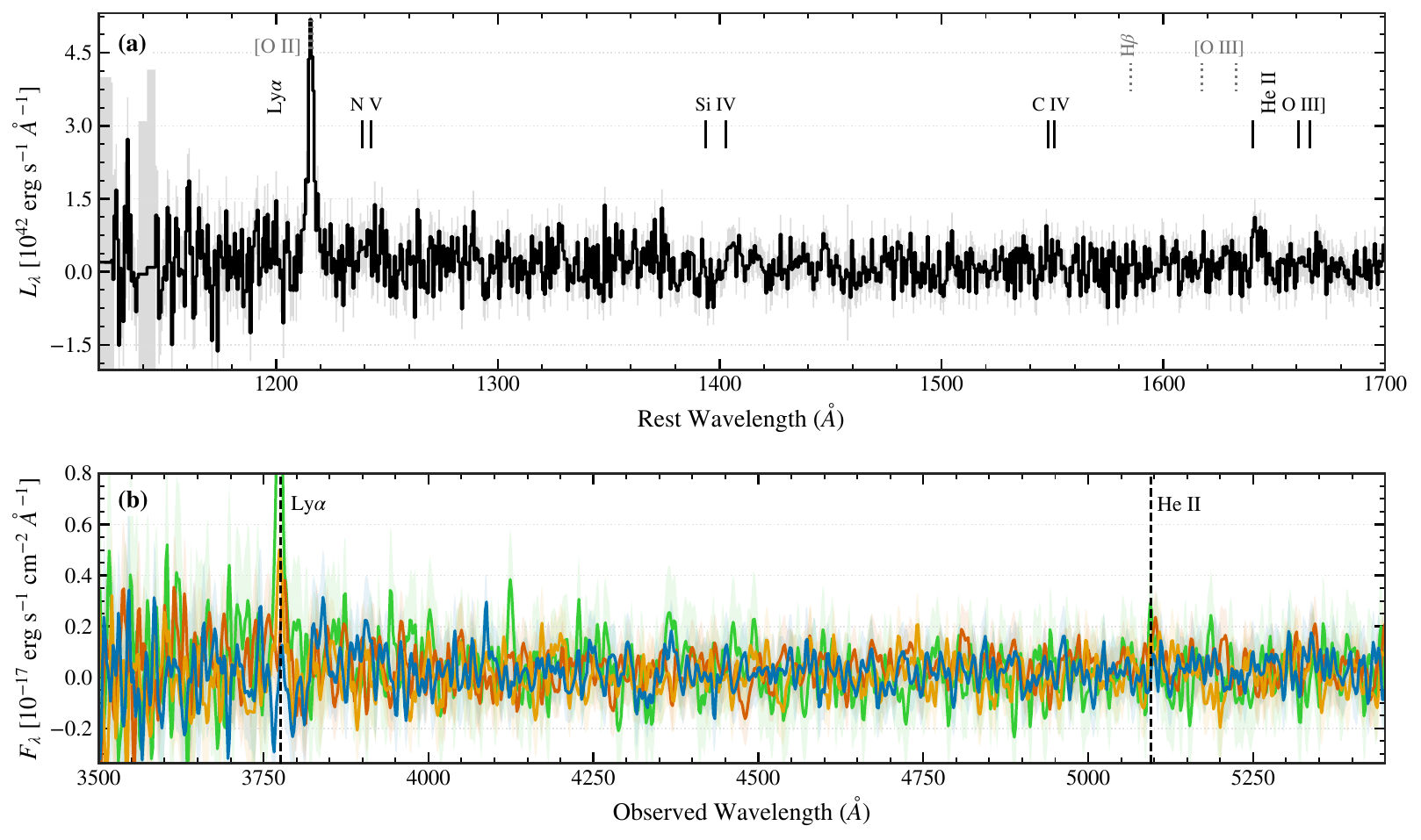}}\\[6pt]
\subfloat[detectid 5001528748.]{%
  \candidatepanel{detectid 5001528748}{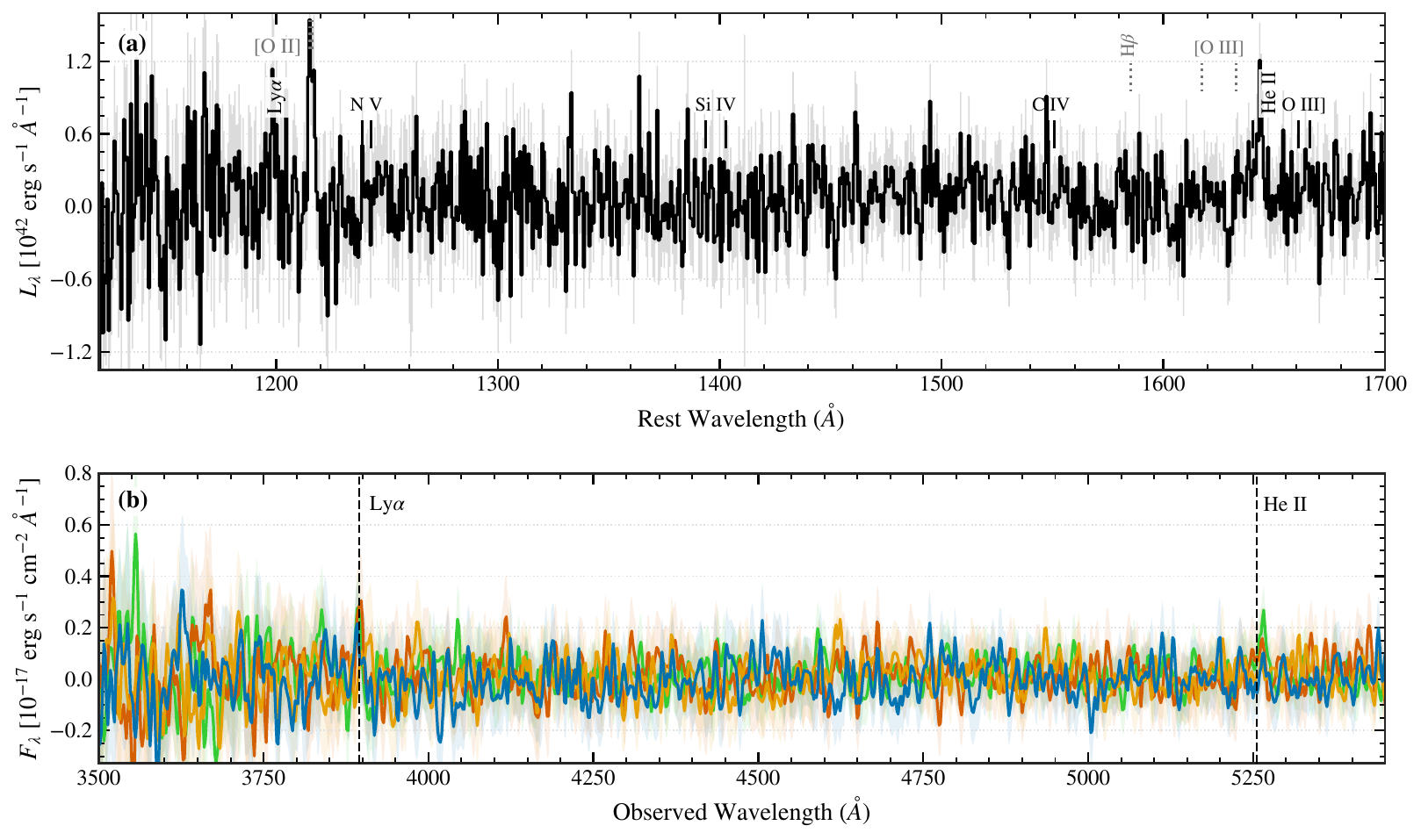}}
\caption{\textbf{Spectral and fiber-level checks for the seven final
candidates.} Continued.}
\end{figure*}

\newcommand{\rejectionpanel}[2]{%
  \IfFileExists{figures/#2}{\includegraphics[width=0.46\textwidth]{figures/#2}}{%
  \fbox{\parbox[c][1.35in][c]{0.43\textwidth}{\centering #1}}}}

\begin{figure*}[t]
\centering
\subfloat[Noise spike.]{\rejectionpanel{Noise-spike rejection}{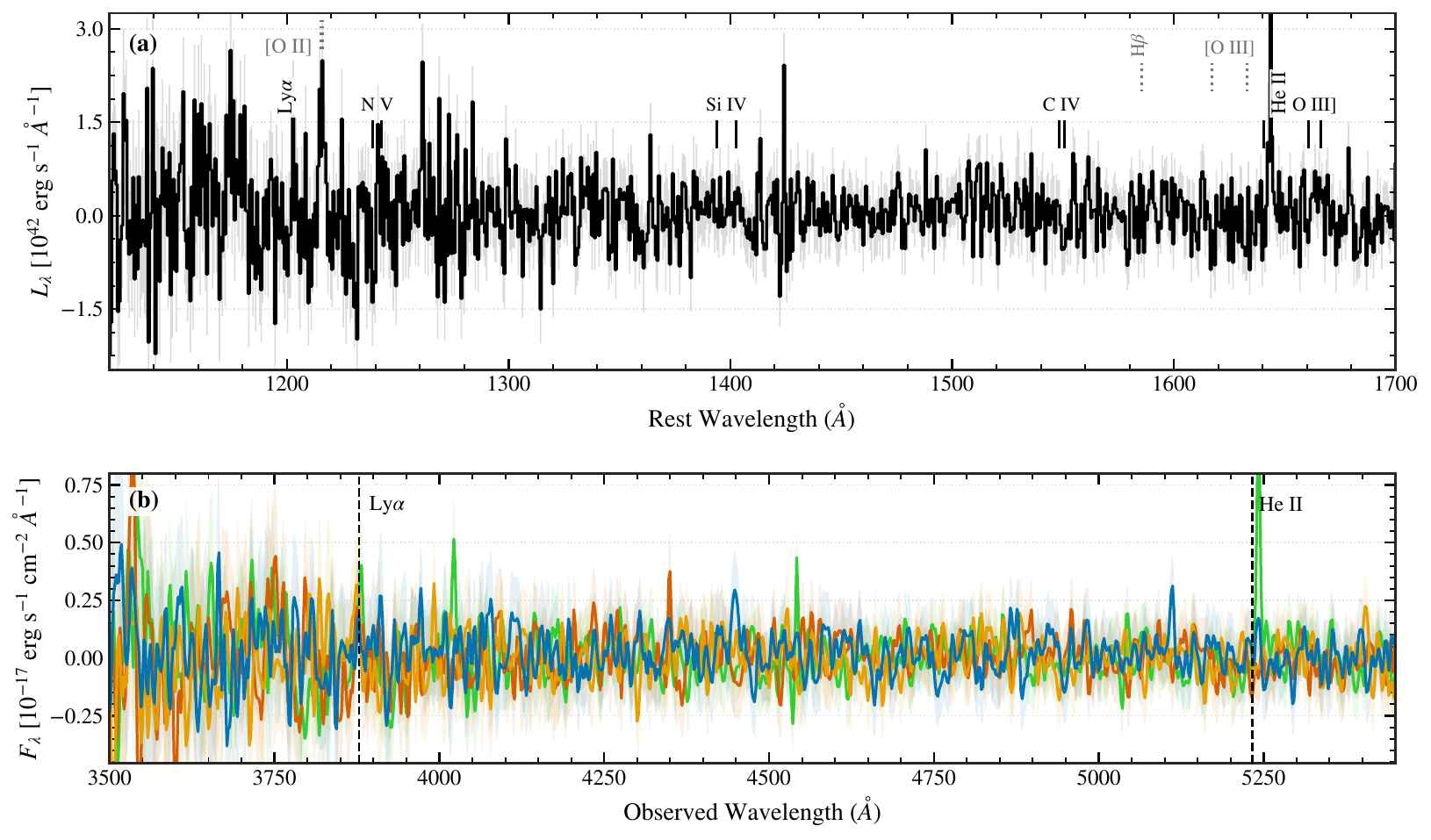}}
\hfill
\subfloat[Double-LAE detection]{\rejectionpanel{Double-LAE rejection}{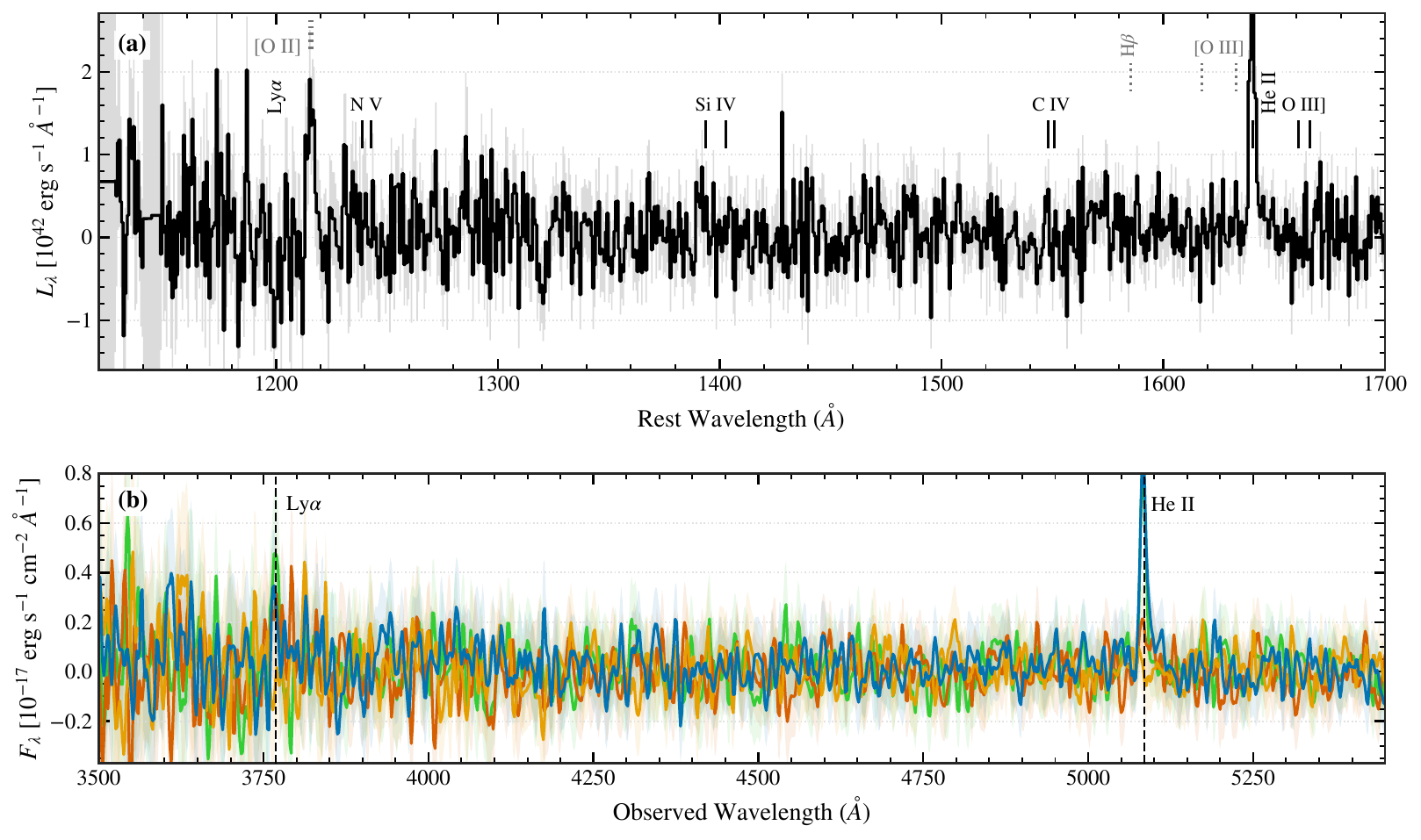}}\\[6pt]
\subfloat[Low-redshift interloper]{\rejectionpanel{Low-redshift interloper}{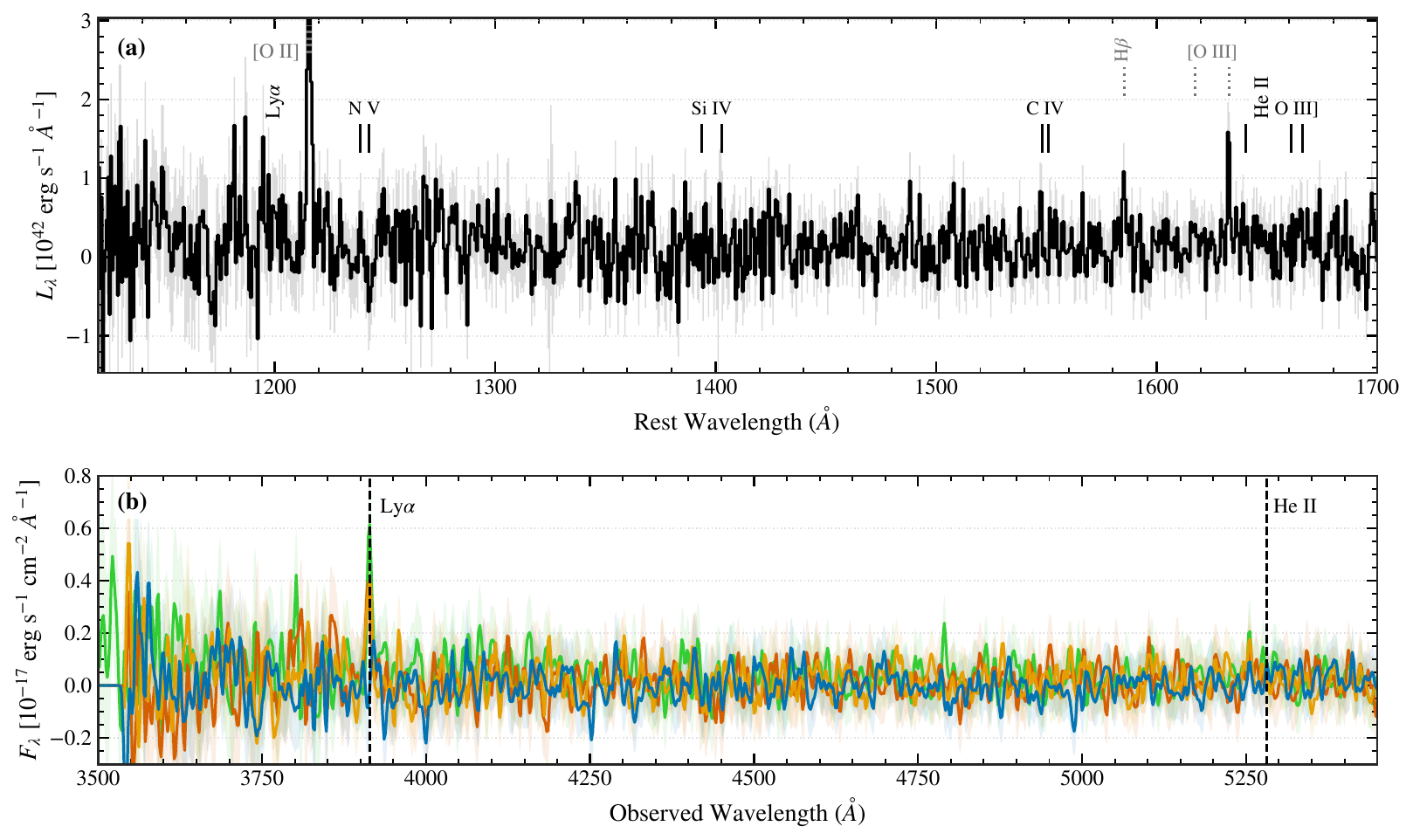}}
\hfill
\subfloat[Metal-enriched source]{\rejectionpanel{Metal-line-rich rejection}{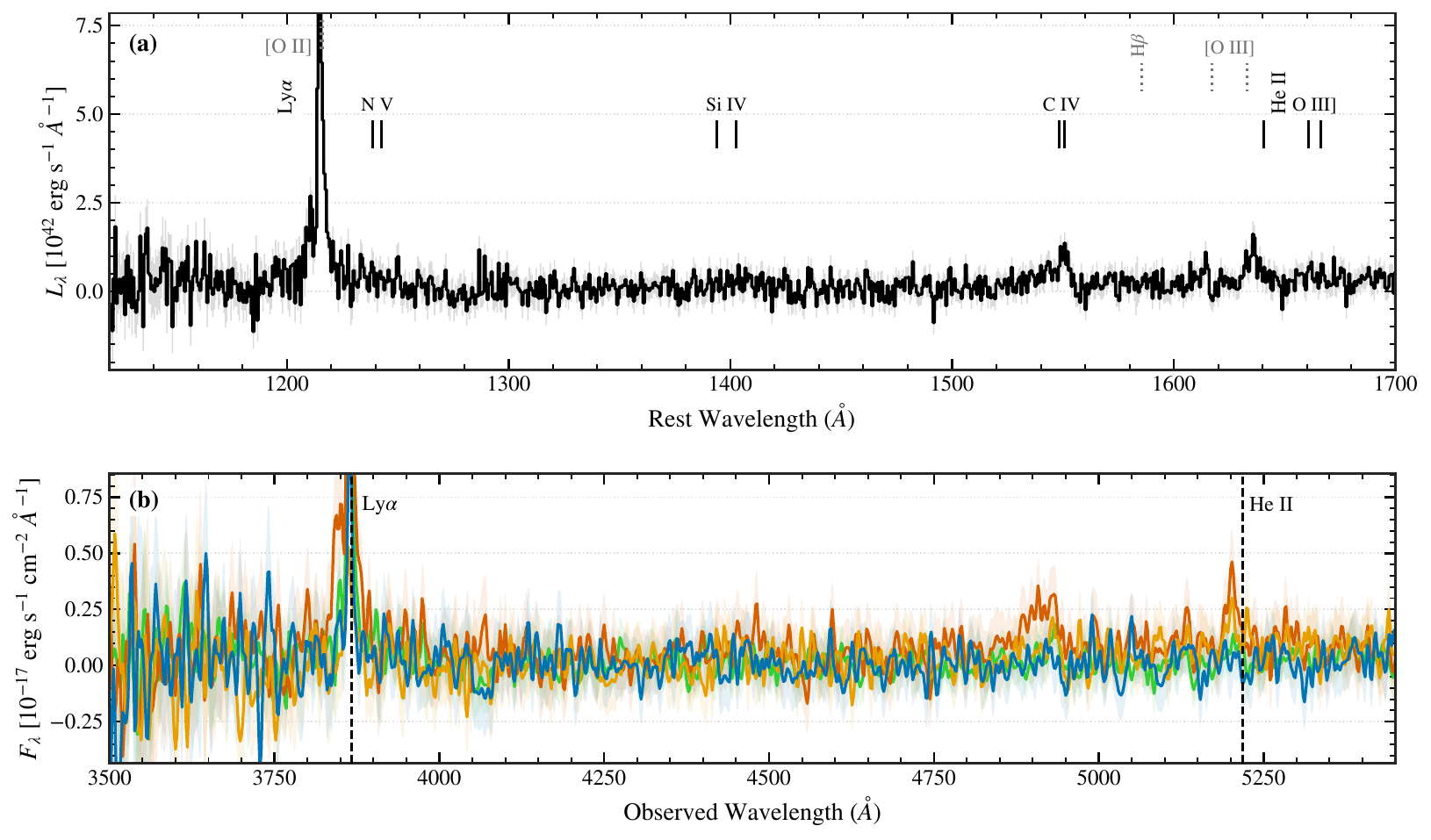}}\\[6pt]
\caption{\textbf{Representative rejected He~\textsc{ii} candidates from the visual-vetting stage}: a noise spike, a double-LAE detection in which the apparent He~\textsc{ii} feature is Ly$\alpha$ from a second source, a low-redshift interloper with an $H\beta$ line, and a metal-enriched source with clear UV metal-line emission.}
\label{fig:rejected_examples}
\end{figure*}

\subsection{Interloper and Ancillary-data Checks}
\label{app:ancillary_interlopers}

We use the available HSC and DECaLS/Legacy Surveys imaging to identify obvious foreground galaxies or blends near each HETDEX extraction. A low-redshift [O~\textsc{ii}] emitter may have a brighter continuum counterpart, extended morphology, or a nearby blended source. The absence of a bright foreground counterpart does not rule out a faint interloper, but it removes the simplest low-redshift explanations for the stacked signal.

Another possible contaminant is a coincident LAE at higher redshift whose Ly$\alpha$ line falls near the expected \heii wavelength. Such a coincidence requires spatial overlap at approximately the $1\farcs5$ VIRUS fiber scale and spectral overlap within the $\sim\pm10$~\AA\ observed-frame centroid window. A rough estimate using the HETDEX LAE surface density and this wavelength interval predicts only $\sim1$ such contaminant in the parent sample. It would also need to be comparable in flux to the primary LAE while remaining undetected in the fiber inspection. These combined requirements make this scenario unlikely to dominate the stack. The highest-ranked source in the inspected sample (fitted \heii S/N = 7.5) provides an example of this failure mode: the apparent \heii feature is \lya emission from a second LAE (Figure~\ref{fig:rejected_examples}).

Where public coverage exists, we inspect X-ray and radio information from eROSITA, Chandra/XMM, LoTSS, and VLASS\null. The absence of a luminous counterpart does not exclude a faint, obscured, or low-luminosity AGN, but it disfavors the simplest luminous-AGN interpretation.

\subsection{Representative Rejected Candidates}
\label{app:rejected_examples}

Figure~\ref{fig:rejected_examples} shows representative failure modes encountered while inspecting the approximately 500 highest-ranked candidates. These examples do not imply that every rejected spectrum falls into one unique category.

\section{Stack Validation and Null Tests}
\label{app:measurement_details}

\subsection{Stack Uncertainty and Robustness Tests}
\label{app:stack_robustness}

For the statistical uncertainty, we combine the MCMC posterior uncertainty with object-level resampling. We generate bootstrap stacks by drawing eight candidates with replacement and recomputing the biweight stack and line measurements. We also perform leave-one-out jackknife tests, recomputing the stack after removing each candidate in turn. The adopted statistical uncertainty conservatively reflects the larger relevant posterior, bootstrap, or jackknife contribution.

We estimate the method-dependent systematic uncertainty from a grid containing biweight and median stacks, narrow and wide continuum windows, Gaussian MCMC fits and fixed-window non-parametric integration, and the comparison between \heii-aligned and \lya-aligned measurements. The fiducial values use the biweight stack, wide continuum window, and Gaussian MCMC fits; the scatter across the robustness grid defines the systematic contribution.

The stack is formed in luminosity density. Non-finite or masked pixels and pixels with suspiciously small uncertainties are excluded or assigned an uncertainty floor before fitting. We do not report EWs or line ratios for individual objects: small shifts in the low-S/N continua produce very large EW changes, whereas the stack improves both line and continuum sensitivity while preserving the common rest-frame line pattern.

The leave-one-out stacks give \heii EW values of $24.7$--$31.3$~\AA, and the Gaussian MCMC measurements remain stable for biweight or median stacking and narrow or wide continuum windows. Fixed-window non-parametric integrations give broadly comparable \heii luminosities but greater continuum sensitivity, so we use them as diagnostics rather than as the fiducial measurements.

\subsection{Selection-aware Null Tests}
\label{app:null_tests}

We perform two sets of null tests to separate generic selection effects from tests specific to the final eight candidates. First, we run null searches on the parent sample by shifting the expected \heii wavelength, shuffling redshifts, searching fake rest-frame wavelengths, and searching negative features. Selecting the strongest null features and aligning them by their fitted centroids can produce an artificial peak even when neither the individual features nor their stack are well described by a Gaussian. This selection-bias effect demonstrates why the automated scan alone is not sufficient proof of an astrophysical signal, and why the visual vetting and fiber-level spatial consistency checks are essential. We therefore do not use the \heii-aligned stack as an independent proof of the false-positive rate.

Second, we apply null tests to the final eight vetted candidates. We stack nearby blank rest-frame windows and large velocity-offset windows from the same eight spectra using the same stacking machinery. These windows do not produce a positive feature comparable to the real \heii stack. Because these spectra already pass the fiber-level consistency checks and full visual vetting, the blank-window result separates the candidate \heii feature from generic stacking artifacts or continuum structure in the final sample.

Together, these tests support a cautious interpretation: the automated scan and centroid-aligned stack alone cannot establish the absolute false-positive rate, while the combination of visual vetting, spatial consistency, final-sample null tests, and weak UV metal-line limits provides the relevant evidence. Deeper spectroscopy is still required to confirm the individual \heii features.

\section{Ly$\alpha$ Measurement and Spatial Tests}
\label{app:lya_tests}

\subsection{Ly$\alpha$--He~\textsc{ii} Kinematics}
\label{app:lya_kinematics}

For each candidate, we refit the \lya and \heii centroids in the PSF-weighted spectrum and use the sign convention defined in Section~\ref{subsec:line_fitting}. Paired draws from the two centroid posteriors are propagated through the velocity-offset definition. For the \heii-aligned stack, the uncertainty additionally includes object bootstrap resampling and the finite precision of the individual \heii centroids used for alignment. The reported uncertainties therefore include the available centroid, profile-choice, bootstrap, and alignment contributions.

The representative source in Figure~\ref{fig:example_candidate_3005020033} has $\Delta v_{\rm Ly\alpha-HeII}=+404^{+86}_{-95}~\kms$. A large positive value can be consistent with a high H~\textsc{i} column, outflow kinematics, covering fraction, or viewing geometry. The present data do not distinguish among these possibilities. Across all eight objects, the mixture of significant positive, significant negative, and near-zero values establishes kinematic diversity and no unique flow direction.

\subsection{Continuum, Profile, and Equivalent-width Robustness}
\label{app:lya_profile_robustness}

The fiducial wide Ly$\alpha$ continuum uses sidebands at 1188--1203 and 1228--1236~\AA. Because the continuum blueward of Ly$\alpha$ can be affected by absorption in the source and along the line of sight, we do not reconstruct the missing flux. The reported Ly$\alpha$ luminosity and EW are observed quantities. We test red side only continua at 1228--1236~\AA\ and at 1228--1236 plus 1268--1295~\AA. We also test several alternative Ly$\alpha$ profile measurements. The split Gaussian uses a common centroid and amplitude but independent Gaussian widths on the blue and red sides of the line. The red-wing-only Gaussian fits a symmetric Gaussian using only wavelengths redward of Ly$\alpha$ rest wavelength and uses the resulting profile to estimate the total line flux. As non-parametric checks, we directly integrate the flux over 1205--1227~\AA\ and also use a red-wing-mirrored measurement, which integrates from the observed Ly$\alpha$ peak to 1227~\AA\ and doubles that flux to estimate the unabsorbed blue-side contribution.

The fiducial MCMC model consists of a symmetric Gaussian emission line plus a local linear continuum. The sampled parameters are the line amplitude, line center, line width, continuum level, continuum slope, and a bounded additional jitter term that allows for underestimation of the pixel uncertainties. We remove invalid pixels and impose a local uncertainty floor before fitting. For the fiducial MCMC measurement, line luminosities, EWs, centroids, and widths are the posterior medians, with intervals from the 16th--84th percentiles.

Replacing the two-sided continuum with either red-side-only continuum changes the symmetric-Gaussian Ly$\alpha$ luminosity by less than $0.1\%$ and the EW by less than $0.4\%$. The split Gaussian changes the luminosity by approximately $2\%$ and the EW by approximately $2.4$~\AA. Across the full parametric and non-parametric grid, the luminosity changes by at most approximately $33\%$, whereas the EW changes much more because the fitted continuum is uncertain. We therefore retain the symmetric Gaussian as the fiducial flux measurement and treat the observed Ly$\alpha$ EW as continuum- and profile-limited.

The measured ${\rm EW}_{\rm Ly\alpha,rest}=61.8\pm59.3$~\AA\ is consequently not a precise discriminator between stellar populations, despite the larger intrinsic Ly$\alpha$ EWs predicted by Pop~III models \citep[e.g.,][]{malhotra_2002,schaerer_2003,raiter_2010}. Resonant scattering, neutral-gas geometry, anisotropic escape, or dust can reduce the emergent EW, whereas ionizing-photon escape can reduce production of recombination photons before Ly$\alpha$ is generated. None of these effects is independently constrained here, and the weak continuum remains the dominant EW limitation.

\subsection{Figure~2 Blue-window Tests}
\label{app:figure2_blue}

For Figure~\ref{fig:example_candidate_3005020033}, the average flux in the far-blue 3500--3650~\AA\ edge window is positive at $2.3\sigma$, consistent with the increased VIRUS noise toward the blue edge (see Figure 1 of \citealt{davis_2023_lae}). The 3670--3820~\AA\ blue-interior window is consistent with zero ($0.37\sigma$). The immediate $-1800$ to $-300~\kms$ interval blueward of Ly$\alpha$ is positive at $5.0\sigma$, whereas the $+1800$ to $+3300~\kms$ red-continuum window is consistent with zero ($0.19\sigma$). These results disfavor a broad additive baseline or common sky residual and instead support localized source-associated Ly$\alpha$ profile structure.

\subsection{PSF-based Ly$\alpha$ Spatial Constraints}
\label{app:lya_spatial}

We quantify the \lya spatial extent using continuum-subtracted line-flux maps and the shot-specific PSF, following the surface-brightness modeling approach of \citet{mentuch_cooper_2026_LAN}. For each candidate, we construct a pseudo-narrowband \lya map by integrating the fiber spectra over the fitted line region and subtracting adjacent continuum. We fit the map with both a PSF-only model and a PSF plus circular exponential component.

We assess spatial extension using the exponential-component fit, the source size relative to the shot-specific PSF, and the improvement of the PSF-plus-exponential model relative to the PSF-only model. We adopt the resolved-source classification defined by \citet{mentuch_cooper_2026_LAN}, which combines the relevant size and model-selection requirements.

We quantify model preference using the corrected Akaike information criterion (AICc) and Bayesian information criterion (BIC). We define $\Delta{\rm AICc}$ and $\Delta{\rm BIC}$ as the PSF-plus-exponential value minus the PSF-only value, so negative values favor the extended model. The fitted exponential scale lengths span $r_s=3.07\pm0.08$ to $17.20\pm0.25$~kpc, but for sources classified as unresolved, these values are only best-fitting model outputs and do not establish real Ly$\alpha$ extension. One candidate, detectid 3007998098, is formally resolved. It has $r_s=17.20\pm0.25$~kpc and $r_{\rm iso}=41.84$~kpc. The extended model is strongly preferred for this source, with $\Delta{\rm AICc}=-54.14$ and $\Delta{\rm BIC}=-33.86$.

The remaining seven candidates do not satisfy the complete resolved-source criterion. Detectids 3010547400 and 4019081078 show suggestive broad profiles and favorable $\Delta{\rm AICc}$ values, but their $\Delta{\rm BIC}$ values are positive, so they remain formally unresolved. The other five sources show weaker or inconsistent evidence for extension. The individual \heii features do not have sufficient S/N for an analogous quantitative size measurement. Table~\ref{tab:lya_spatial} summarizes the \lya spatial fit diagnostics for all eight candidates.

\begin{deluxetable}{crrrrc}
\tabletypesize{\scriptsize}
\tablecaption{Ly$\alpha$ Spatial Diagnostics}
\label{tab:lya_spatial}
\tablehead{
\colhead{detectid} &
\colhead{$r_s$ (kpc)} &
\colhead{$r_{\rm iso}$ (kpc)} &
\colhead{$\Delta{\rm AICc}$} &
\colhead{$\Delta{\rm BIC}$} &
\colhead{Resolved?}
}
\startdata
3005020033          & $10.53\pm0.32$ & 19.42 &  $+0.52$ & $+19.26$ & No  \\
{3007998098} & ${17.20\pm0.25}$ & {41.84} & ${-54.14}$ & ${-33.86}$ & {Yes} \\
3009751099          & $ 3.07\pm0.08$ & 13.95 &  $+5.95$ & $+26.00$ & No  \\
3010547400          & $15.58\pm0.04$ & 33.96 & $-16.37$ &  $+3.64$ & No  \\
4019081078          & $17.10\pm0.18$ & 27.75 & $-19.48$ &  $+0.80$ & No  \\
4028177445          & $ 8.49\pm0.45$ & 18.17 &  $+5.65$ & $+24.80$ & No  \\
4028632963          & $ 9.86\pm0.26$ & 27.56 &  $+0.73$ & $+19.88$ & No  \\
5001528748          & $15.90\pm0.11$ & 16.95 &  $-1.07$ & $+19.22$ & No
\enddata
\tablecomments{
The exponential scale length is denoted by $r_s$, and $r_{\rm iso}$ is the maximum radius at which the Ly$\alpha$ surface brightness remains above the adopted isophotal threshold (see text in Appendix \ref{app:lya_spatial}). The resolved classification follows the criteria of \citet{mentuch_cooper_2026_LAN}.
}
\end{deluxetable}

\end{document}